\documentclass{article}
\usepackage{amsmath,amsthm,upref}
\usepackage{amssymb,amsbsy}
\usepackage{amsfonts}
\usepackage{cite}
\newtheorem{example}{Example}

\usepackage{mathrsfs,bm}
\renewcommand{\mathcal}{\mathscr}
\renewcommand{\mathbf}{\bm}

\normalfont\upshape
\numberwithin{equation}{section}

\title{\textbf{Generalized duality, Hamiltonian formalism}\\
 \textbf{and new brackets}}
\author{\emph{Steven Duplij}\footnote{\emph{On leave of absence from}: V.N. Karazin
Kharkov National University, Svoboda Sq. 4, Kharkov 61022, Ukraine, sduplij@gmail.com}\\[5pt]
\\Center for Mathematics, Science and Education,\\ 
Rutgers University,
118 Frelinghuysen Rd., \\
Piscataway, NJ 08854, USA\\[5pt] 
\small duplij@math.rutgers.edu,
http://math.rutgers.edu/\~{}duplij}

\begin{document}

\maketitle

\begin{abstract}
It is shown that any singular Lagrangian theory: 1) can be formulated  
 without the use of constraints 
 by introducing a Clairaut-type version of the Hamiltonian formalism;
 2) leads to a special kind of nonabelian gauge theory which is similar to the Poisson gauge theory; 
 3) can be treated as the many-time classical dynamics. 
A generalization of the Legendre transform to the 
zero Hessian case is done by using the mixed (envelope/general) solution of 
the multidimensional Clairaut equation. The corresponding system of motion equations 
 is equivalent to the Lagrange equations and 
has a linear algebraic subsystem 
for ``unresolved'' velocities. 
Then the equations of motion are written in the Hamilton-like
 form by introducing new antisymmetric brackets. 
 This is a ``shortened'' formalism, since it
 does not contain the ``nondynamical''  (degenerate) momenta at all, and therefore 
 there is no notion of constraint. 
 It is outlined that 
any classical degenerate Lagrangian theory 
(in its Clairaut-type Hamiltonian form) is equivalent to the many-time 
classical dynamics. Finally, the relation between the presented formalism 
and the Dirac approach to constrained systems is given.

\end{abstract}

\medskip

\tableofcontents


\bigskip

\section{Introduction}

Nowadays, many fundamental physical models are based on gauge field theories
\cite{weinberg123,del/eti/fre}. On the classical level, they are described by
singular (degenerate) Lagrangians, which makes the passage to the Hamiltonian
description, which is important for quantization, highly nontrivial and
complicated (see, e.g., \cite{sundermeyer,reg/tei}).

A common way to deal with singular theories is the Dirac approach \cite{dirac}
based on extending the phase space and constraints. This treatment of
constrained theories has been deeply reviewed, e.g., in lecture notes
\cite{wipf} and books \cite{git/tyu,hen/tei}. In spite of its general success,
the Dirac approach has some problems \cite{dis83,pon3,wan/ran} and is not
directly applicable in some cases, e.g., for irregular constrained systems
(with linearly dependent constraints) \cite{mik/zan,bal06} or so-called
\textquotedblleft pathological examples\textquotedblright\ \cite{lus91}.
Therefore, it is worthwhile to reconsider basic ideas of the Hamiltonian
formalism in general from another point of view \cite{dup_belg2009,dup11}.

In the standard approach for nonsingular theories \cite{ste,arnold1}, the
transition from Lagrangian to Hamiltonian description is done by the Legendre
transform, and then finding the Hamiltonian as an envelope solution of the
corresponding Clairaut equation \cite{kam,izu94}. The main idea of our
formulation is the following \cite{dup2011}: for singular theories, instead of
the Lagrange multiplier procedure developed by Dirac \cite{dirac}, we
construct and solve the corresponding multidimensional Clairaut equation
\cite{izu94}. In this way, we state that the ordinary duality can be
generalized to the Clairaut duality \cite{dup2011}.

In this paper we develop our previous work \cite{dup_belg2009,dup2011} to
construct a self-consistent\ analog of the canonical (Hamiltonian) formalism
and present a general algorithm to describe any Lagrangian system (singular or
not) as a set of first-order differential equations without introducing
Lagrange multipliers. From mathematical viewpoint, we extend to the singular
dynamical systems the well-known construction of Hamiltonian as a solution of
the Clairaut equation, which was developed in \cite{arnold1} for unconstrained
systems. To simplify matters, we consider systems with a finite number of
degrees of freedom, use local coordinates and the clear language of
differential equations together with the Clairaut equation theory
\cite{kam,izu94}.

Using the fact that for a singular Lagrangian system the Hessian matrix is
degenerate and therefore has the rank which is less than its size, we separate
the dynamical variables into \textquotedblleft physical\textquotedblright\ (or
regular/non-degenerate) and \textquotedblleft non-physical\textquotedblright%
\ (degenerate) ones, such that the Hessian matrix of the former is
non-degenerate. On the other hand the Clairaut equation has two kinds of
solutions: the general one and the envelope one \cite{izu94}. The key idea is
to use the envelope solution for the \textquotedblleft
physical\textquotedblright\ variables and the general solution for
\textquotedblleft non-physical\textquotedblright\ ones, and therefore the
separation of variables is unavoidable. In this way we obtain a unique analog
of the Hamiltonian (called the mixed Hamilton-Clairaut function) which
(formally) coincides with one derived by the geometric approach
\cite{car6,tul} and by the generalized Legendre transformation
\cite{cen/hol/hoy/mar}. Then, using the mixed Hamilton-Clairaut function we
make a passage from the second order Lagrange equations of motion to a set of
the first order Hamilton-like equations. The next important step is exclusion
of the so-called degenerate \textquotedblleft momenta\textquotedblright\ and
introducing the \textquotedblleft physical\textquotedblright%
\ Hamilton-Clairaut function (which corresponds to the total Dirac
Hamiltonian), which allows us to present the equations of motion as a system
of differential equations for \textquotedblleft physical\textquotedblright%
\ coordinates and momenta together with a system of linear equations for
unresolved (\textquotedblleft non-physical\textquotedblright) velocities. The
different kinds of the solutions of this system of linear equations leads to
the classification of singular systems, which reminds the classification of
constraints, but does not coincide with it: the former does not contain
analogs of higher constraints, because there are no corresponding degenerate
\textquotedblleft momenta\textquotedblright\ at all. Some formulations without
(primary) constraints were treated in \cite{git/tyu1,der2005,malt78}, and
without any constraints, but for special (regularizable) kind of Lagrangians,
see \cite{kru/sme,kru02}.

The \textquotedblleft shortened\textquotedblright\ approach can play an
important role for quantization of such complicated constrained systems as
gauge field theories \cite{khv} and gravity \cite{ran08}. To illustrate the
power and simplicity of our method, we consider such examples, as the Cawley
Lagrangian \cite{caw79} (which leads to difficulties in the Dirac approach),
and the relativistic particle. A novel Hamilton-like form of the equations of
motion is achieved by defining new antisymmetric brackets. The quantization of
such brackets can be done by means of the standard methods (see, e.g.,
\cite{gre/rei}) without the using of the Dirac quantization \cite{dirac}.

While analyzing the part of the equations of motion corresponding to
\textquotedblleft unresolved\textquotedblright\ velocities, we arrive
effectively at a kind of nonabelian gauge theory in the \textquotedblleft
degenerate\textquotedblright\ coordinate subspace, which is similar to the
Poisson gauge theory \cite{flo/ili/tik}. But in our case the partial
derivatives and Poisson brackets \textquotedblleft live\textquotedblright\ in
different subspaces. We also outline that the Clairaut-type formulation is
equivalent to the many-time classical dynamics developed in
\cite{dom/lon/gom/pon,lon/lus/pon}, if the \textquotedblleft
nondynamical\textquotedblright\ (degenerate) coordinates are treated as the
additional \textquotedblleft times\textquotedblright. Finally, in Appendix we
show that, after introduction of \textquotedblleft
non-dynamical\textquotedblright\ momenta, corresponding Lagrange multipliers
and respective constraints, the Clairaut-type formulation presented here
corresponds to the Dirac approach \cite{dirac}.

\section{The Legendre-Fenchel and Legendre transforms}

We start with a brief description of the standard Legendre-Fenchel and
Legendre transforms for the theory with nondegenerate Lagrangian
\cite{arnold,rockafellar}. Let $L\left(  q^{A},v^{A}\right)  $, $A=1,\ldots
n$, be a Lagrangian given by a function of $2n$ variables ($n$ generalized
coordinates $q^{A}$ and $n$ velocities $v^{A}=\dot{q}^{A}=dq^{A}/dt$) on the
configuration space $\mathsf{T}M$, where $M$ is a smooth manifold. We use
indices in arguments, because we need to distinguish different kinds of
coordinates (similar to \cite{tul/urb}). For the same reason, we use the
summation signs with explicit ranges. Also, we consider the time-independent
case for simplicity and conciseness, which will not influence on the main procedure.

By the convex approach definition (see e.g. \cite{roc67,arnold}), a
Hamiltonian $H\left(  q^{A},p_{A}\right)  $ is a dual function on the phase
space $\mathsf{T}^{\ast}M$ (or convex conjugate \cite{rockafellar}) to the
Lagrangian (in the second set of variables $p_{A}$) and is constructed by
means of the \textit{Legendre-Fenchel transform} $L\overset{\mathfrak{Leg}%
^{\mathrm{Fen}}}{\longmapsto}H^{\text{\textrm{Fen}}}$ defined by
\cite{fen49,roc67}%
\begin{align}
H^{\text{$\mathrm{Fen}$}}\left(  q^{A},p_{A}\right)   &  =\sup\limits_{v^{A}%
}G\left(  q^{A},v^{A},p_{A}\right)  ,\label{hqp}\\
G\left(  q^{A},v^{A},p_{A}\right)   &  =\sum_{B=1}^{n}p_{B}v^{B}-L\left(
q^{A},v^{A}\right)  . \label{g}%
\end{align}

Note that this definition is very general, and it can be applied to nonconvex
\cite{ala/mai/roc} and nondifferentiable \cite{val/hji/sav} functions
$L\left(  q^{A},v^{A}\right)  $, which can lead to numerous extended versions
of the Hamiltonian formalism (see, e.g., \cite{cla77,roc96,iof}). Also, a
generalization of the convex conjugacy can be achieved by substituting in
(\ref{g}) the form $p_{A}v^{A}$ by any function $\Psi\left(  p_{A}%
,v^{A}\right)  $ satisfying special conditions \cite{goe/roc}.

In the standard mechanics \cite{goldstein1}, one usually restricts to convex,
smooth and differentiable Lagrangians (see, e.g., \cite{arnold,sud/muk}). Then
the coordinates $q^{A}(t)$ are treated as fixed (passive with respect to the
Legendre transform) parameters, and the velocities $v^{A}(t)$ are assumed
independent functions of time.

According to our assumptions the supremum (\ref{hqp}) is attained by finding
an extremum point $v^{A}=v_{extr}^{A}$ of the (\textquotedblleft
pre-Hamiltonian\textquotedblright) function $G\left(  q^{A},v^{A}%
,p_{A}\right)  $, which leads to the supremum condition
\begin{equation}
p_{B}=\left.  \dfrac{\partial L\left(  q^{A},v^{A}\right)  }{\partial v^{B}%
}\right\vert _{v^{A}=v_{extr}^{A}}. \label{pl}%
\end{equation}

It is commonly assumed (see, e.g., \cite{arnold,sud/muk,goldstein1}) that the
only way to get rid of dependence on the velocities $v^{A}$ in the r.h.s. of
(\ref{hqp}) is to resolve (\ref{pl}) with respect to velocities and find its
solution given by a set of functions
\begin{align}
v_{extr}^{B}=V^{B}\left(  q^{A},p_{A}\right)  .
\end{align}
This can be done only in the class of nondegenerate Lagrangians $L\left(
q^{A},v^{A}\right)  =L^{\text{\textrm{nondeg}}}\left(  q^{A},v^{A}\right)  $
(in the second set of variables $v^{A}$), which is equivalent to the Hessian
being non-zero%
\begin{equation}
\det\left\Vert \dfrac{\partial^{2}L^{\text{\textrm{nondeg}}}\left(
q^{A},v^{A}\right)  }{\partial v^{B}\partial v^{C}}\right\Vert \neq0.
\label{hs}%
\end{equation}
Then substitute $v_{extr}^{A}$ to (\ref{hqp}) and obtain the standard
Hamiltonian (see, e.g., \cite{arnold,goldstein1})%
\begin{align}
&  H\left(  q^{A},p_{A}\right)  \overset{def}{=}G\left(  q^{A},v_{extr}%
^{A},p_{A}\right) \nonumber\\
&  = \sum_{B=1}^{n}p_{B}V^{B}\left(  q^{A},p_{A}\right)
-L^{\text{\textrm{nondeg}}}\left(  q^{A},V^{A}\left(  q^{A},p_{A}\right)
\right)  . \label{hq}%
\end{align}

The passage from the nondegenerate Lagrangian $L^{\text{\textrm{nondeg}}%
}\left(  q^{A},v^{A}\right)  $ to the Hamiltonian $H\left(  q^{A}%
,p_{A}\right)  $ is called the \textit{Legendre transform} (of functions)
which will be denoted by $L^{\text{\textrm{nondeg}}}\overset{\mathfrak{Leg}%
}{\longmapsto}H$.

\ In the geometric approach \cite{tulczyjew,mar/sal/sim/vit,abr/mar}, the
\textit{Legendre transform} of functions $L^{\text{\textrm{nondeg}}}%
\overset{\mathfrak{Leg}}{\longmapsto}H$ is tantamount to the \textit{Legendre
transformation} from the configuration space to the phase space $\mathsf{Leg}%
:\mathsf{T}M\rightarrow\mathsf{T}^{\ast}M$ (or between submanifolds in the
presence of constraints \cite{men/tul,bat/gom/pon/rom,gia/man/sar}).
Nevertheless, here we will use local coordinates and the language of
differential equations which are responsible to corresponding function
transforms, in particular the Clairaut equation theory \cite{kam,izu94}.

\section{The Legendre-Clairaut transform}

The connection between the Legendre transform, convexity and the Clairaut
equation has a long story \cite{kam,ste} (see also \cite{arnold1}). Here we
present an alternative way to deal with the supremum condition (\ref{pl}) and
consider the related multidimensional Clairaut equation (which was proposed in
\cite{dup_belg2009}).

We differentiate (\ref{hq}) by the momenta $p_{A}$ and use the supremum
condition (\ref{pl}) to get
\begin{align}
&  \dfrac{\partial H\left(  q^{A},p_{A}\right)  }{\partial p_{B}}=V^{B}\left(
q^{A},p_{A}\right) \nonumber\\
&  + \sum_{C=1}^{n}\left(  p_{C}-\left.  \dfrac{\partial L\left(  q^{A}%
,v^{A}\right)  }{\partial v^{C}}\right\vert _{v^{C}=V^{C}\left(  q^{A}%
,p_{A}\right)  }\right)  \dfrac{\partial V^{C}\left(  q^{A},p_{A}\right)
}{\partial p_{B}}=V^{B}\left(  q^{A},p_{A}\right)  , \label{hv}%
\end{align}
which can be called the \textit{dual supremum condition} (indeed this gives
the first set of the Hamilton equations, see below). The relations (\ref{pl}),
(\ref{hq}) and (\ref{hv}) together represent a particular case of the Donkin
theorem (see e.g. \cite{goldstein1}).

Then we substitute (\ref{hv}) in (\ref{hq}) and obtain%
\begin{equation}
H\left(  q^{A},p_{A}\right)  \equiv\sum_{B=1}^{n}p_{B}\dfrac{\partial H\left(
q^{A},p_{A}\right)  }{\partial p_{B}}-L^{\mathrm{nondeg}}\left(  q^{A}%
,\dfrac{\partial H\left(  q^{A},p_{A}\right)  }{\partial p_{C}}\right)  ,
\label{cl0}%
\end{equation}
which contains no manifest dependence on velocities at all. It is important
that, for nonsingular Lagrangians, the relation (\ref{cl0}) is an
\textit{identity}, which follows from (\ref{pl}), (\ref{hq}) and (\ref{hv}) by
our construction. This relation can be also obtained from the geometric
approach \cite{car/mar/sim/vit/zac}.

Now we make the main step: to treat the equation (\ref{cl0}) \textit{by
itself} (without referring to (\ref{pl}), (\ref{hq}) and (\ref{hv})) as a
\textit{definition of a new transform} being a solution of the following
nonlinear partial differential equation (the multidimensional Clairaut
equation) \cite{dup_belg2009,dup2011}
\begin{equation}
H^{\text{$\mathrm{Cl}$}}\left(  q^{A},\lambda_{A}\right)  =\sum_{B=1}%
^{n}\lambda_{B}\dfrac{\partial H^{\text{$\mathrm{Cl}$}}\left(  q^{A}%
,\lambda_{A}\right)  }{\partial\lambda_{B}}-L\left(  q^{A},\dfrac{\partial
H^{\text{$\mathrm{Cl}$}}\left(  q^{A},\lambda_{A}\right)  }{\partial
\lambda_{A}}\right)  , \label{cl}%
\end{equation}
in the formal independent variables $\lambda_{A}$ (initially not connected
with $p_{A}$ defined by (\ref{pl})) and $L\left(  q^{A},v_{A}\right)  $ is
\textit{any} differentiable smooth function of $2n$ variables $q^{A},v_{A}$,
where coordinates $q^{A}$ play the role of external parameters. It is very
important that here in (\ref{cl}) we do not demand the nondegeneracy condition
(\ref{hs}) imposed on $L\left(  q^{A},v_{A}\right)  $. We call the transform
defined by the equation (\ref{cl}) $L\overset{\mathfrak{Leg}%
^{\text{\textrm{Cl}}}}{\longmapsto}H^{\text{\textrm{Cl}}}$ a \textit{Clairaut
duality transform} (or the \textit{Legendre-Clairaut transform}) and
$H^{\text{\textrm{Cl}}}\left(  q^{A},\lambda_{A}\right)  $ a
\textit{Hamilton-Clairaut function} \cite{dup_belg2009,dup2011}.

Note that the relation (\ref{pl}) which is commonly treated as a definition of
all dynamical momenta $p_{A}$, in our approach is the supremum condition for
some of the independent variables of the Clairaut duality transform
$\lambda_{A}$. In the differential equation language, $\lambda_{A}$ are
independent mathematical variables having no connection with any physical
dynamics. Before solving the Clairaut equation (\ref{cl}) and applying the
supremum condition (\ref{pl}) which is in our language is $\lambda_{A}%
=p_{A}=\partial L\diagup\partial v^{A}$, the independent variables
$\lambda_{A}$ are not connected with the Lagrangian and cannot be called
momenta. The role of $\lambda_{A}$ is to find all possible solutions of the
Clairaut equation (\ref{cl}) for nondegenerate and degenerate Lagrangians
$L\left(  q^{A},v^{A}\right)  $ . Only those of $\lambda_{A}$ which will be
restricted by the supremum condition (\ref{pl}) can be interpreted as momenta
with the corresponding geometric description in terms of the cotangent space.

The difference between the Legendre-Clairaut transform and the Legendre
transform is crucial for degenerate Lagrangian theories \cite{dup_belg2009}.
Specifically, the multidimensional Clairaut equation (\ref{cl}) has solutions
even for degenerate Lagrangians $L\left(  q^{A},v^{A}\right)  =L^{\mathrm{deg}%
}\left(  q^{A},v^{A}\right)  $ when the Hessian is zero%
\begin{equation}
\det\left\Vert \dfrac{\partial^{2}L^{\mathrm{deg}}\left(  q^{A},v^{A}\right)
}{\partial v^{B}\partial v^{C}}\right\Vert =0. \label{hs1}%
\end{equation}

In this case, the Legendre-Clairaut transform of functions (\ref{cl})
$\mathfrak{Leg}^{\text{\textrm{Cl}}}$ is another (along with the
Legendre-Fenchel transform $\mathfrak{Leg}^{\mathrm{Fen}}$) counterpart to the
ordinary Legendre transform (\ref{hq}) in the case of degenerate Lagrangians.
The Clairaut equation (\ref{cl}) has solutions always, which is independent of
resolving the supremum condition (\ref{pl}) with respect to velocities and
properties of the Hessian.

To make this manifest and to find solutions of the Clairaut equation
(\ref{cl}), we differentiate it by $\lambda_{C}$ to obtain%
\begin{equation}
\sum_{B=1}^{n}\left[  \lambda_{B}-\left.  \dfrac{\partial L\left(  q^{A}%
,v^{A}\right)  }{\partial v^{B}}\right\vert _{v^{B}=\tfrac{\partial
H^{\text{\textrm{Cl}}}\left(  q^{A},\lambda_{A}\right)  }{\partial\lambda_{B}%
}}\right]  \cdot\dfrac{\partial^{2}H^{\text{\textrm{Cl}}}\left(  q^{A}%
,\lambda_{A}\right)  }{\partial\lambda_{B}\partial\lambda_{C}}=0. \label{plh}%
\end{equation}
Now we apply the ordinary method of solving the Clairaut equation (see
Appendix \ref{sec-clair}). There are two possible solutions of (\ref{plh}),
one in which the square brackets vanish (envelope solution) and one in which
the double derivative in velocity vanishes (general solution). The l.h.s. of
(\ref{plh}) is a sum over $B$ and it is quite conceivable that one may vanish
for some $B$ and the other vanish for other $B$. The physical reason of
choosing the particular solution is presented in Section \ref{sec-mixed}.
Thus, we have two solutions of the Clairaut equation:

1) The \textit{envelope solution} defined by the first multiplier in
(\ref{plh}) being zero%
\begin{equation}
\lambda_{B}=p_{B}=\dfrac{\partial L\left(  q^{A},v^{A}\right)  }{\partial
v^{B}}, \label{pb}%
\end{equation}
which coincides with the supremum condition (\ref{pl}), together with
(\ref{hv}). In this way, we obtain the standard Hamiltonian (\ref{hq})%
\begin{equation}
H_{env}^{\text{\textrm{Cl}}}\left(  q^{A},\lambda_{A}\right)  |_{\lambda
_{A}=p_{A}}=H\left(  q^{A},p_{A}\right)  . \label{hh}%
\end{equation}
Thus, in the nondegenerate case, the \textquotedblleft
envelope\textquotedblright\ Legendre-Clairaut transform $\mathfrak{Leg}%
_{env}^{\text{\textrm{Cl}}}:L\rightarrow H_{env}^{\text{\textrm{Cl}}}$
coincides with the ordinary Legendre transform by our construction here.

2) A \textit{general solution} is defined by%
\begin{equation}
\dfrac{\partial^{2}H^{\text{\textrm{Cl}}}\left(  q^{A},\lambda_{A}\right)
}{\partial\lambda_{B}\partial\lambda_{C}}=0 \label{h0}%
\end{equation}
which gives $\dfrac{\partial H^{\text{\textrm{Cl}}}\left(  q^{A},\lambda
_{A}\right)  }{\partial\lambda_{B}}=c^{B}$, here $c^{B}$ are arbitrary smooth
functions of $q^{A}$, and the latter are considered in the Clairaut equation
(\ref{cl}) as parameters (passive variables). Then the general solution
acquires the form%
\begin{equation}
H_{gen}^{\text{\textrm{Cl}}}\left(  q^{A},\lambda_{A},c^{A}\right)
=\sum_{B=1}^{n}\lambda_{B}c^{B}-L\left(  q^{A},c^{A}\right)  , \label{gen}%
\end{equation}
which corresponds to a \textquotedblleft general\textquotedblright%
\ Legendre-Clairaut transform $\mathfrak{Leg}_{gen}^{\text{\textrm{Cl}}%
}:L\rightarrow H_{gen}^{\text{\textrm{Cl}}}$. Note that the general solution
$H_{gen}^{\text{\textrm{Cl}}}\left(  q^{A},\lambda_{A},c^{A}\right)  $ is
always linear in the variables $\lambda_{A}$ and the latter are not actually
the dynamical momenta $p_{A}$, because we do not have the envelope solution
condition (\ref{pb}), and therefore now there is no supremum condition
(\ref{pl}). The variables $c^{A}$ are in fact unresolved velocities $v^{A}$ in
the case of the general solution.

Note that in the standard way, $\mathfrak{Leg}_{env}^{\text{\textrm{Cl}}}$ can
be also obtained by finding the envelope of the general solution
\cite{arnold1}, i.e. differentiating (\ref{gen}) by $c^{A}$ as%
\begin{equation}
\dfrac{\partial H_{gen}^{\text{\textrm{Cl}}}\left(  q^{A},\lambda_{A}%
,c^{A}\right)  }{\partial c^{B}}=\lambda_{B}-\dfrac{\partial L\left(
q^{A},c^{A}\right)  }{\partial c^{B}}=0 \label{hc}%
\end{equation}
which coincides with (\ref{pb}) and (\ref{pl}). This means that $H_{gen}%
^{\text{\textrm{Cl}}}\left(  q^{A},\lambda_{A},c^{A}\right)  |_{c^{A}=v^{A}}$
is in fact the \textquotedblleft pre-Hamiltonian\textquotedblright\ (\ref{g}),
which was needed to find the supremum in (\ref{hqp}).

Let us consider the classical example of one-dimensional oscillator.

\begin{example}
Let $L\left(  x,v\right)  =mv^{2}/2-kx^{2}/2$ ($m$, $k$ are constants), then
the corresponding Clairaut equation (\ref{cl}) for $H=H^{\mathrm{Cl}}\left(
x,\lambda\right)  $ is%
\begin{equation}
H=\lambda H_{\lambda}^{\prime}-\dfrac{m}{2}\left(  H_{\lambda}^{\prime
}\right)  ^{2}+\dfrac{kx^{2}}{2}, \label{hhp}%
\end{equation}
where prime denotes partial differentiation with respect to $\lambda$. The
general solution of (\ref{hhp}) is%
\begin{equation}
H_{gen}^{\text{\textrm{Cl}}}\left(  x,\lambda,c\right)  =\lambda
c-\dfrac{mc^{2}}{2}+\dfrac{kx^{2}}{2}, \label{ho}%
\end{equation}
where $c$ is an arbitrary function (\textquotedblleft unresolved
velocity\textquotedblright\ $v$). The envelope solution (with $\lambda=p$) can
be found from the condition%
\[
\dfrac{\partial H^{\text{\textrm{Cl}}}}{\partial c}=p-mc=0\Longrightarrow
c_{extr}=\dfrac{p}{m},
\]
which gives%
\begin{equation}
H_{env}^{\text{\textrm{Cl}}}\left(  x,p\right)  =\dfrac{p^{2}}{2m}%
+\dfrac{kx^{2}}{2} \label{he}%
\end{equation}
in the standard way.
\end{example}

\begin{example}
Let $L\left(  x,v\right)  =x\exp kv$, then the corresponding Clairaut equation
for $H=H^{\mathrm{Cl}}\left(  x,\lambda\right)  $ is%
\begin{equation}
H=\lambda H_{\lambda}^{\prime}-x\exp\left(  kH_{\lambda}^{\prime}\right)  .
\end{equation}
The general solution is%
\begin{equation}
H_{gen}^{\text{\textrm{Cl}}}\left(  x,\lambda\right)  =\lambda c-x\exp kc,
\label{hg}%
\end{equation}
where $c$ is any smooth function of $x$.

The envelope solution (with $\lambda=p$) can be found by differentiating the
general solution (\ref{hg})%
\[
\dfrac{\partial H^{\text{\textrm{Cl}}}}{\partial c}=p-x\exp
kc=0\Longrightarrow c_{extr}=\dfrac{1}{k}\ln\dfrac{p}{x},
\]
which leads to%
\begin{equation}
H_{env}^{\text{\textrm{Cl}}}\left(  x,p\right)  =\dfrac{p}{k}\ln\dfrac{p}%
{x}-p.
\end{equation}

\end{example}

\section{The mixed\ Legendre-Clairaut transform\label{sec-mixed}}

Now consider a singular Lagrangian $L\left(  q^{A},v^{A}\right)
=L^{\mathrm{deg}}\left(  q^{A},v^{A}\right)  $ for which the Hessian is zero
(\ref{hs1}). This means that the rank of the Hessian matrix $W_{AB}%
=\tfrac{\partial^{2}L\left(  q^{A},v^{A}\right)  }{\partial v^{B}\partial
v^{C}}$ is $r<n$, and we suppose that $r$ is constant. We rearrange indices of
$W_{AB}$ in such a way that a nonsingular minor of rank $r$ appears in the
upper left corner \cite{git/tyu0}. Represent the index $A$ as follows: if
$A=1,\ldots,r$, we replace $A$ with $i$ (the \textquotedblleft
regular\textquotedblright\ index), and, if $A=r+1,\ldots,n$ we replace $A$
with $\alpha$ (the \textquotedblleft degenerate\textquotedblright\ index).
Obviously, $\det W_{ij}\neq0$, and $\operatorname{rank}W_{ij}=r$. Thus any set
of variables labelled by a single index splits as a disjoint union of two
subsets. We call those subsets \textit{regular} (having Latin indices) and
\textit{degenerate} (having Greek indices).

The standard Legendre transform $\mathfrak{Leg}$ is not applicable in the
degenerate case because the condition (\ref{hs}) is not valid \cite{car6,tul}.
Therefore the supremum condition (\ref{pl}) cannot be resolved with respect to
degenerate $A$, but it can be resolved only for regular $A$, because $\det
W_{ij}\neq0$. On the contrary, the Clairaut duality transform given by
(\ref{cl}) is independent of the Hessian being zero or not \cite{dup_belg2009}%
. Thus, we state the main idea of the formalism we present here: \textit{the
ordinary duality can be generalized to the Clairaut duality}. This can be
rephrased by saying that the standard Legendre transform $\mathfrak{Leg}$
(given by (\ref{hq})) can be generalized to the singular Lagrangian theory
using the Legendre-Clairaut transform $\mathfrak{Leg}^{\text{\textrm{Cl}}}$
given by the multidimensional Clairaut equation (\ref{cl}).

To find its solutions, we differentiate (\ref{cl}) by $\lambda_{A}$ and split
the sum (\ref{plh}) in $B$ as follows%
\begin{align}
&  \sum_{i=1}^{r}\left[  \lambda_{i}-\dfrac{\partial L\left(  q^{A}%
,v^{A}\right)  }{\partial v^{i}}\right]  \cdot\dfrac{\partial^{2}%
H^{\text{\textrm{Cl}}}\left(  q^{A},\lambda_{A}\right)  }{\partial\lambda
_{i}\partial\lambda_{C}}\nonumber\\
&  +\sum_{\alpha=r+1}^{n}\left[  \lambda_{\alpha}-\dfrac{\partial L\left(
q^{A},v^{A}\right)  }{\partial v^{\alpha}}\right]  \cdot\dfrac{\partial
^{2}H^{\text{\textrm{Cl}}}\left(  q^{A},\lambda_{A}\right)  }{\partial
\lambda_{\alpha}\partial\lambda_{C}}=0. \label{pphh}%
\end{align}

As $\det W_{ij}\neq0$, we suggest to replace (\ref{pphh}) by the conditions%
\begin{align}
&  \lambda_{i}=p_{i}=\dfrac{\partial L\left(  q^{A},v^{A}\right)  }{\partial
v^{i}},\ \ i=1,\ldots,r,\label{pp}\\
&  \dfrac{\partial^{2}H^{\text{\textrm{Cl}}}\left(  q^{A},\lambda_{A}\right)
}{\partial\lambda_{\alpha}\partial\lambda_{C}}=0,\ \ \ \alpha=r+1,\ldots n.
\label{hpp}%
\end{align}

In this way we obtain a \textit{mixed\ envelope/general solution} of the
Clairaut equation as follows \cite{dup_belg2009}. We resolve (\ref{pp}) by
regular velocities $v^{i}=V^{i}\left(  q^{A},p_{i},c^{\alpha}\right)  $ and
write down a solution of (\ref{hpp}) as
\begin{equation}
\dfrac{\partial H^{\text{\textrm{Cl}}}\left(  q^{A},\lambda_{A}\right)
}{\partial\lambda_{\alpha}}=c^{\alpha}, \label{hcc}%
\end{equation}
where $c^{\alpha}$ are arbitrary variables being the unresolved velocities
$v^{\alpha}$. In this way we obtain a \textit{mixed\ Hamilton-Clairaut}
function%
\begin{align}
H_{mix}^{\text{\textrm{Cl}}}\left(  q^{A},p_{i},\lambda_{\alpha},v^{\alpha
}\right)   &  =\sum_{i=1}^{r}p_{i}V^{i}\left(  q^{A},p_{i},v^{\alpha}\right)
\nonumber\\
&  +\sum_{\beta=r+1}^{n}\lambda_{\beta}v^{\beta}-L\left(  q^{A},V^{i}\left(
q^{A},p_{i},v^{\alpha}\right)  ,v^{\alpha}\right)  , \label{hm}%
\end{align}
which is the desired \textquotedblleft mixed\textquotedblright%
\ Legendre-Clairaut transform of functions $L\overset{\mathfrak{Leg}%
_{mix}^{\text{\textrm{Cl}}}}{\longmapsto}H_{mix}^{\text{\textrm{Cl}}}$ written
in coordinates.

Note that (\ref{hm}) was obtained formally as a mixed general/envelope
solution of the Clairaut equation for the sought-for Hamilton-Clairaut
function without any reference to the dynamics (this connection will be
considered in next section). Nevertheless, $H_{mix}^{\text{\textrm{Cl}}}$
coincides with the corresponding functions derived from the \textquotedblleft
slow and careful Legendre transformation\textquotedblright\ \cite{tul/urb} and
with the \textquotedblleft generalized Legendre
transformation\textquotedblright\ \cite{cen/hol/hoy/mar}, as well as from the
implicit partial differential equation on the cotangent bundle
\cite{mar91,daz} in the local coordinates \cite{yos/mar} and in the general
geometric approach \cite{mar/men/tul}.

\begin{example}
Let $L\left(  x,y,v_{x},v_{y}\right)  =myv_{x}^{2}/2+kxv_{y}$, then the
corresponding Clairaut equation for $H=H^{\mathrm{Cl}}\left(  x,y,\lambda
_{x},\lambda_{y}\right)  $ is%
\begin{equation}
H=\lambda_{x}H_{\lambda_{x}}^{\prime}+\lambda_{y}H_{\lambda_{y}}^{\prime
}-\dfrac{my}{2}\left(  H_{\lambda_{x}}^{\prime}\right)  ^{2}-kxH_{\lambda_{y}%
}^{\prime}. \label{hp}%
\end{equation}
The general solution of (\ref{hp}) is%
\[
H_{gen}^{\text{\textrm{Cl}}}\left(  x,y,\lambda_{x},\lambda_{y},c_{x}%
,c_{y}\right)  =\lambda_{x}c_{x}+\lambda_{y}c_{y}-\dfrac{myc_{x}^{2}}%
{2}-kxc_{y},
\]
where $c_{x}$, $c_{y}$ are arbitrary functions of the passive variables $x,y$.
Then we differentiate%
\begin{align*}
\dfrac{\partial H_{gen}^{\text{\textrm{Cl}}}}{\partial c_{x}}  &
=p_{x}-myc_{x}=0,\ \ \Longrightarrow c_{x}^{extr}=\dfrac{p_{x}}{my},\\
\dfrac{\partial H_{gen}^{\text{\textrm{Cl}}}}{\partial c_{y}}  &  =\lambda
_{y}-kx.
\end{align*}
Finally, we solve the first equation with respect to $c_{x}$ and set
$c_{y}\longmapsto v_{y}$ an \textquotedblleft unresolved
velocity\textquotedblright\ and obtain the mixed Hamiltonian-Clairaut function%
\begin{equation}
H_{mix}^{\text{\textrm{Cl}}}\left(  x,y,p_{x},\lambda_{y},v_{y}\right)
=\dfrac{p_{x}^{2}}{2my}+v_{y}\left(  \lambda_{y}-kx\right)  . \label{hpx}%
\end{equation}

This result can be compared with one obtained in the geometric approach by
reduction of the Hamiltonian Morse family in \cite{tul/urb}.
\end{example}

\section{Hamiltonian formulation of singular Lagrangian systems}

Let us use the mixed\ Hamilton-Clairaut function $H_{mix}^{\text{\textrm{Cl}}%
}\left(  q^{A},p_{i},\lambda_{\alpha},v^{\alpha}\right)  $ (\ref{hm}) to
describe a singular Lagrangian theory by a system of ordinary first-order
differential equations. In our formulation we divide the set of standard
Lagrange equations of motion
\begin{equation}
\frac{d}{dt}\frac{\partial L\left(  q^{A},v^{A}\right)  }{\partial v^{B}%
}=\frac{\partial L\left(  q^{A},v^{A}\right)  }{\partial q^{B}} \label{leq}%
\end{equation}
into two subsets, according to the index $B$ being regular ($B=i=1,\ldots,r$)
or degenerate ($B=\alpha=r+1,\ldots n$). We use the designation
of\ \textquotedblleft physical\textquotedblright\ momenta (\ref{pp}) in the
regular subset only, such that the Lagrange equations become%
\begin{align}
\dfrac{dp_{i}}{dt}  &  =\dfrac{\partial L\left(  q^{A},v^{A}\right)
}{\partial q^{i}},\label{ph}\\
\dfrac{dB_{\alpha}\left(  q^{A},p_{i}\right)  }{dt}  &  =\left.
\dfrac{\partial L\left(  q^{A},v^{A}\right)  }{\partial q^{\alpha}}\right\vert
_{v^{i}=V^{i}\left(  q^{A},p_{i},v^{\alpha}\right)  }, \label{ph1}%
\end{align}
where%
\begin{equation}
B_{\alpha}\left(  q^{A},p_{i}\right)  \overset{def}{=}\left.  \dfrac{\partial
L\left(  q^{A},v^{A}\right)  }{\partial v^{\alpha}}\right\vert _{v^{i}%
=V^{i}\left(  q^{A},p_{i},v^{\alpha}\right)  } \label{h}%
\end{equation}
are given functions which determine dynamics of the singular Lagrangian system
in the \textquotedblleft degenerate\textquotedblright\ sector. The functions
$B_{\alpha}\left(  q^{A},p_{i}\right)  $ are independent of the unresolved
velocities $v^{\alpha}$ since $\operatorname{rank}W_{AB}=r$. One should also
take into account that now
\begin{equation}
\frac{dq^{i}}{dt}=V^{i}\left(  q^{A},p_{i},v^{\alpha}\right)  ,\;\;\,\;
\frac{dq^{\alpha}}{dt}=v^{\alpha}.
\end{equation}
Note that before imposing the Lagrange equations (\ref{ph}) (while solving the
Clairaut equation (\ref{cl})), the arguments of $L\left(  q^{A},v^{A}\right)
$ were treated as independent variables.

A passage to an analog of the Hamiltonian formalism can be done by the
standard procedure: consider the full differential of both sides of (\ref{hm})
and use the supremum condition (\ref{pp}), which gives (note that in previous
sections the Lagrange equations of motion (\ref{leq}) were not used)
\begin{align*}
\dfrac{\partial H_{mix}^{\mathrm{Cl}}}{\partial p_{i}}  &  =V^{i}\left(
q^{A},p_{i},v^{\alpha}\right)  ,\\
\dfrac{\partial H_{mix}^{\mathrm{Cl}}}{\partial\lambda_{\alpha}}  &
=v^{\alpha},\\
\dfrac{\partial H_{mix}^{\mathrm{Cl}}}{\partial q^{i}}  &  =-\left.
\dfrac{\partial L\left(  q^{A},v^{A}\right)  }{\partial q^{i}}\right\vert
_{v^{i}=V^{i}\left(  q^{A},p_{i},v^{\alpha}\right)  }+\sum_{\beta=r+1}%
^{n}\left[  \lambda_{\beta}-B_{\beta}\left(  q^{A},p_{i}\right)  \right]
\dfrac{\partial v^{\beta}}{\partial q^{i}},\\
\dfrac{\partial H_{mix}^{\mathrm{Cl}}}{\partial q^{\alpha}}  &  =-\left.
\dfrac{\partial L\left(  q^{A},v^{A}\right)  }{\partial q^{\alpha}}\right\vert
_{v^{i}=V^{i}\left(  q^{A},p_{i},v^{\alpha}\right)  }+\sum_{\beta=r+1}%
^{n}\left[  \lambda_{\beta}-B_{\beta}\left(  q^{A},p_{i}\right)  \right]
\dfrac{\partial v^{\beta}}{\partial q^{\alpha}}.
\end{align*}
An application of (\ref{ph}) yields the system of equations which gives a
Hamiltonian-Clairaut description of a singular Lagrangian system%
\begin{align}
\dfrac{\partial H_{mix}^{\mathrm{Cl}}}{\partial p_{i}}  &  =\dfrac{dq^{i}}%
{dt},\label{h1}\\
\dfrac{\partial H_{mix}^{\mathrm{Cl}}}{\partial\lambda_{\alpha}}  &
=\dfrac{dq^{\alpha}}{dt},\label{h2}\\
\dfrac{\partial H_{mix}^{\mathrm{Cl}}}{\partial q^{i}}  &  =-\dfrac{dp_{i}%
}{dt}+\sum_{\beta=r+1}^{n}\left[  \lambda_{\beta}-B_{\beta}\left(  q^{A}%
,p_{i}\right)  \right]  \dfrac{\partial v^{\beta}}{\partial q^{i}}%
,\label{h3}\\
\dfrac{\partial H_{mix}^{\mathrm{Cl}}}{\partial q^{\alpha}}  &  =\dfrac
{dB_{\alpha}\left(  q^{A},p_{i}\right)  }{dt}+\sum_{\beta=r+1}^{n}\left[
\lambda_{\beta}-B_{\beta}\left(  q^{A},p_{i}\right)  \right]  \dfrac{\partial
v^{\beta}}{\partial q^{\alpha}}. \label{h4}%
\end{align}

The system (\ref{h1})--(\ref{h4}) has two disadvantages: 1) it contains the
\textquotedblleft nondynamical\ momenta\textquotedblright\ $\lambda_{\alpha}$;
2) it has derivatives of unresolved velocities $v^{\alpha}$. We observe that
we can get rid of these difficulties, if we reformulate (\ref{h1})--(\ref{h4})
by introducing a \textquotedblleft physical\textquotedblright\ Hamiltonian%
\begin{equation}
H_{phys}\left(  q^{A},p_{i}\right)  =H_{mix}^{\mathrm{Cl}}\left(  q^{A}%
,p_{i},\lambda_{\alpha},v^{\alpha}\right)  -\sum_{\beta=r+1}^{n}\left[
\lambda_{\beta}-B_{\beta}\left(  q^{A},p_{i}\right)  \right]  v^{\beta},
\label{hph}%
\end{equation}
which does not depend on the variables $\lambda_{\alpha}$ (\textquotedblleft
nondynamical\ momenta\textquotedblright) at all by construction%
\begin{equation}
\dfrac{\partial H_{phys}}{\partial\lambda_{\alpha}}=0 \label{hl}%
\end{equation}
(cf. (\ref{hcc}) and (\ref{hm})). Then the \textquotedblleft
physical\textquotedblright\ Hamiltonian (\ref{hph}) can be rewritten in the
form%
\begin{align}
&  H_{phys}\left(  q^{A},p_{i}\right)  =\sum_{i=1}^{r}p_{i}V^{i}\left(
q^{A},p_{i},v^{\alpha}\right) \nonumber\\
&  +\sum_{\alpha=r+1}^{n}B_{\alpha}\left(  q^{A},p_{i}\right)  v^{\alpha
}-L\left(  q^{A},V^{i}\left(  q^{A},p_{i},v^{\alpha}\right)  ,v^{\alpha
}\right)  . \label{hph1}%
\end{align}
Using (\ref{pp}), we can show that the r.h.s. of (\ref{hph1}) indeed does not
depend on $\lambda_{\alpha}$ and degenerate velocities $v^{\alpha}$%
\begin{equation}
\dfrac{\partial H_{phys}}{\partial v^{\alpha}}=0, \label{hpv}%
\end{equation}
which justifies the term \textquotedblleft physical\textquotedblright.
Therefore, the time evolution of the singular Lagrangian system (\ref{leq}) is
determined by $\left(  n-r+1\right)  $ functions $H_{phys}\equiv
H_{phys}\left(  q^{A},p_{i}\right)  $ and $B_{\alpha}\equiv B_{\alpha}\left(
q^{A},p_{i}\right)  $. Writing $\left(  q^{A},p_{i}\right)  =\left(
q^{\alpha}|q^{i},p_{i}\right)  \in R^{n-r}\times Sp\left(  r,r\right)  \equiv
M_{phys}$, where $R^{n-r}$ is a real space of dimension $\left(  n-r\right)  $
and $Sp\left(  r,r\right)  $ is the symplectic space of dimension $\left(
r,r\right)  $, we observe that $H_{phys}:R^{n-r}\times Sp\left(  r,r\right)
\rightarrow R$ and $B_{\alpha}:R^{n-r}\times Sp\left(  r,r\right)  \rightarrow
R^{n-r}$.

Then we obtain from (\ref{h1})--(\ref{h4}) the main result of our
Clairaut-type formulation: the sought-for system of ordinary first-order
differential equations (\textit{the Hamilton-Clairaut system}) which describes
any singular Lagrangian classical system (satisfying the second order Lagrange
equations (\ref{leq})) has the form%
\begin{align}
&  \dfrac{dq^{i}}{dt}=\left\{  q^{i},H_{phys}\right\}  _{phys}-\sum
_{\beta=r+1}^{n}\left\{  q^{i},B_{\beta}\right\}  _{phys}\dfrac{dq^{\beta}%
}{dt},\ \ i=1,\ldots r\label{q1}\\
&  \dfrac{dp_{i}}{dt}=\left\{  p_{i},H_{phys}\right\}  _{phys}-\sum
_{\beta=r+1}^{n}\left\{  p_{i},B_{\beta}\right\}  _{phys}\dfrac{dq^{\beta}%
}{dt},\ \ i=1,\ldots r\label{q2}\\
&  \sum_{\beta=r+1}^{n}\left[  \dfrac{\partial B_{\beta}}{\partial q^{\alpha}%
}-\dfrac{\partial B_{\alpha}}{\partial q^{\beta}}+\left\{  B_{\alpha}%
,B_{\beta}\right\}  _{phys}\right]  \dfrac{dq^{\beta}}{dt}\nonumber\\
&  =\dfrac{\partial H_{phys}}{\partial q^{\alpha}}+\left\{  B_{\alpha
},H_{phys}\right\}  _{phys},\ \ \ \ \ \ \ \ \ \ \ \ \alpha=r+1,\ldots,n
\label{q3}%
\end{align}
where%
\begin{equation}
\left\{  X,Y\right\}  _{phys}=\sum_{i=1}^{n-r}\left(  \frac{\partial
X}{\partial q^{i}}\frac{\partial Y}{\partial p_{i}}-\frac{\partial Y}{\partial
q^{i}}\frac{\partial X}{\partial p_{i}}\right)  \label{xyp}%
\end{equation}
is the \textquotedblleft physical\textquotedblright\ Poisson bracket (in
regular variables $q^{i}$, $p_{i}$) for functions $X$ and $Y$ on $M_{phys}$.

The system (\ref{q1})--(\ref{q3}) is equivalent to the Lagrange equations of
motion (\ref{leq}) by construction. Thus, the Clairaut-type formulation
(\ref{q1})--(\ref{q3}) is valid for any Lagrangian theory without additional
conditions, as opposite to other approaches (see, e.g. \cite{pon3,wan/ran}).

\begin{example}
\textrm{(Cawley \cite{caw79})} Let $L=\dot{x}\dot{y}+zy^{2}/2$, then the
equations of motion are%
\begin{equation}
\ddot{x}=yz,\ \ \ \ddot{y}=0,\ \ \ y^{2}=0. \label{xyz}%
\end{equation}
Because the Hessian has rank 2, and the velocity $\dot{z}$ does not enter into
the Lagrangian, the only degenerate velocity is $\dot{z}$ ($\alpha=z)$, the
regular momenta are $p_{x}=\dot{y}$, $p_{y}=\dot{x}$ ($i=x,y$). Thus, we have%
\[
H_{phys}=p_{x}p_{y}-\dfrac{1}{2}zy^{2},\ \ \ \ B_{z}=0.
\]
The equations of motion (\ref{q1})--(\ref{q2}) are%
\begin{equation}
\dot{p}_{x}=0,\ \ \ \dot{p}_{y}=yz, \label{pyz}%
\end{equation}
and the condition (\ref{q3}) gives%
\begin{equation}
\dfrac{\partial H_{phys}}{\partial z}=-\dfrac{1}{2}y^{2}=0. \label{hy}%
\end{equation}
Observe that (\ref{pyz}) and (\ref{hy}) coincide with the initial Lagrange
equations of motion (\ref{xyz}).
\end{example}

Note, that because the number of equations $r+r+n-r=n+r$ coincides with the
number of the sought-for variables $n_{q^{i}}=r$, $n_{p_{i}}=r$,
$n_{q^{\alpha}}=n-r$, there are no constraints in (\ref{q1})--(\ref{q3}) at
all. In particular, the system (\ref{q3}) has the number $\left(  n-r\right)
$ equations, which exactly coincides with the number of the sought-for
\textquotedblleft unresolved\textquotedblright\ velocities $v^{\alpha}%
=\tfrac{dq^{\alpha}}{dt}$. Therefore (\ref{q3}) is a standard system of linear
algebraic equations with respect to $v^{\alpha}$, but not constraints (when
there are more sought-for variables, than equations).

\begin{example}
\textrm{(\cite{bat/sni,sch98})} Let us consider a classical particle on
$R^{3}$ with the regular Lagrangian $\left(  \dot{x}^{2}+\dot{y}^{2}+\dot
{z}^{2}\right)  /2$ subject to the nonholonomic constraint $\dot{z}=y\dot{x}$.
To apply the Clairaut equation method, we introduce an extra coordinate $u$,
then this system is equivalent to the singular Lagrangian system on $R^{4}$
described by%
\begin{equation}
L=\dfrac{\dot{x}^{2}+\dot{y}^{2}+\dot{z}^{2}}{2}+u\left(  \dot{z}-y\dot
{x}\right)  . \label{lx}%
\end{equation}
The Lagrange equations of motion are straightforward (cf. \cite{bat/sni})%
\begin{equation}
\ddot{x}-\dot{u}y-\dot{y}u=0,\ \ \ddot{y}+u\dot{x}=0,\ \ \ \ddot{z}+\dot
{u}=0,\ \ \dot{z}-y\dot{x}=0. \label{xx}%
\end{equation}
The Hessian of (\ref{lx}) is zero, and so the system is singular. Because the
rank of the Hessian matrix $\mathrm{diag}\left(  1,1,1,0\right)  $ is 3, we
have 3 regular and 1 degenerate variables. First, we should find the
\textquotedblleft physical\textquotedblright\ Hamiltonian using the Clairaut
equation formalism, then we make passage from the second order equations
(\ref{xx}) to the first order equations similar to (\ref{q1})--(\ref{q3}). Let
us consider the multidimensional Clairaut equation (\ref{cl}) for the
Hamilton-Clairaut function $H\equiv H^{Cl}\left(  x,y,z,u,\lambda_{x}%
,\lambda_{y},\lambda_{z},\lambda_{u}\right)  $%
\begin{align}
H  &  =\lambda_{x}H_{\lambda_{x}}^{\prime}+\lambda_{y}H_{\lambda_{y}}^{\prime
}+\lambda_{z}H_{\lambda_{z}}^{\prime}+\lambda_{u}H_{\lambda_{u}}^{\prime
}\nonumber\\
&  -\dfrac{1}{2}\left(  H_{\lambda_{x}}^{\prime}\right)  ^{2}-\dfrac{1}%
{2}\left(  H_{\lambda_{y}}^{\prime}\right)  ^{2}-\dfrac{1}{2}\left(
H_{\lambda_{z}}^{\prime}\right)  ^{2}-uH_{\lambda_{z}}^{\prime}+yuH_{\lambda
_{x}}^{\prime}. \label{hxu}%
\end{align}
The general solution of (\ref{hxu}) is%
\begin{equation}
H_{gen}=\lambda_{x}c_{x}+\lambda_{y}c_{y}+\lambda_{z}c_{z}+\lambda_{u}%
c_{u}-\dfrac{c_{x}^{2}+c_{y}^{2}+c_{z}^{2}}{2}-uc_{z}+yuc_{x}, \label{hlc}%
\end{equation}
where initially $c_{x},c_{y},c_{z},c_{u}$ are arbitrary functions of the
passive (with respect to the Clairaut equation (\ref{hxu}) variables
$x,y,z,u$. To find the supremum conditions (\ref{hc}), we write derivatives%
\begin{align}
\dfrac{\partial H_{gen}}{\partial c_{x}}  &  =\lambda_{x}-c_{x}+yu=0,\\
\dfrac{\partial H_{gen}}{\partial c_{y}}  &  =\lambda_{y}-c_{y}=0,\\
\dfrac{\partial H_{gen}}{\partial c_{z}}  &  =\lambda_{z}-c_{z}-u=0,\\
\dfrac{\partial H_{gen}}{\partial c_{u}}  &  =\lambda_{u}.
\end{align}
Observe, that only 3 first conditions here can be resolved with respect to
$c_{i}$ ($i=x,y,z$), and therefore indeed these $\lambda_{i}$ correspond to
the \textquotedblleft physical\textquotedblright\ momenta (\ref{pp}), that is
$\lambda_{i}=p_{i}=\partial L\diagup\partial v_{i}$ ($i=x,y,z$). So the
extremum values of $c_{i}$ are%
\begin{equation}
c_{x}^{extr}=p_{x}+yu,\ \ \ \ c_{y}^{extr}=p_{y},\ \ \ \ c_{z}^{extr}=p_{z}-u,
\label{cc}%
\end{equation}
while $c_{u}$ becomes the \textquotedblleft unresolved\textquotedblright%
\ velocity $c_{u}=v_{u}$. In this way, inserting (\ref{cc}) into (\ref{hlc}),
for the mixed Hamilton-Clairaut function (\ref{hm}) we have%
\begin{equation}
H_{mix}^{Cl}=\dfrac{p_{x}^{2}+p_{y}^{2}+p_{z}^{2}}{2}+\lambda_{u}%
v_{u}+u\left(  yp_{x}-p_{z}\right)  +u^{2}\dfrac{1+y^{2}}{2}.
\end{equation}
Now we calculate the function (\ref{h}) and the \textquotedblleft
physical\textquotedblright\ Hamiltonian (\ref{hph}) as%
\begin{align}
B_{u}  &  =\dfrac{\partial L}{\partial\dot{u}}=0,\\
H_{phys}  &  =\dfrac{p_{x}^{2}+p_{y}^{2}+p_{z}^{2}}{2}+u\left(  yp_{x}%
-p_{z}\right)  +u^{2}\dfrac{1+y^{2}}{2},
\end{align}
which indeed does not depend on $\lambda_{u}$ and $v_{u}$. Thus, using
(\ref{q1})--(\ref{q3}), we obtain the Hamilton-Clairaut system%
\begin{align}
&  \dot{x}=p_{x}+yu,\ \ \ \dot{y}=p_{y},\ \ \ \dot{z}=p_{z}-u,\label{yp1}\\
&  \dot{p}_{x}=0,\ \ \ \dot{p}_{y}=-u\left(  p_{x}+yu\right)  ,\ \ \dot{p}%
_{z}=0,\label{yp2}\\
&  yp_{x}-p_{z}+u\left(  1+y^{2}\right)  =0, \label{yp3}%
\end{align}
which coincides with the system of the Lagrange equations of motion (\ref{xx})
by construction. It is remarkable that the \textquotedblleft
degenerate\textquotedblright variable $u$ is determined by the algebraic
equation (\ref{yp3})%
\begin{equation}
u=\dfrac{p_{z}-yp_{x}}{1+y^{2}},
\end{equation}
and therefore the singular system (\ref{lx}) has no \textquotedblleft
gauge\textquotedblright\ degrees of freedom.
\end{example}

In general, if a dynamical system is nonsingular, it has no \textquotedblleft
degenerate\textquotedblright\ variables at all, because the rank of the
Hessian $r$ is full ($r=n$). The distinguishing property of any singular
system ($r<n$) is clear and simple in our Clairaut-type approach: it contains
the additional system of linear algebraic equations (\ref{q3}) for
\textquotedblleft unresolved\textquotedblright\ velocities $v^{\alpha}$ (not
constraints), which can be analyzed and solved by standard linear algebra
methods. Indeed, the linear algebraic system (\ref{q3}) gives a full
classification of singular Lagrangian theories, which is done in the next section.

\begin{example}
The classical relativistic particle is described by%
\begin{equation}
L=-mR,\ \ \ \ R=\sqrt{\dot{x}_{0}^{2}-\sum_{i=x,y,z}\dot{x}_{i}^{2}},
\label{lp}%
\end{equation}
where a dot denotes derivative with respect to the proper time. Because the
rank of the Hessian is 3, we consider the velocities $\dot{x}_{i}$ as regular
variables and the velocity $\dot{x}_{0}$ as a degenerate variable. Then for
the regular canonical momenta we have $p_{i}=\partial L\diagup\partial\dot
{x}_{i}=m\dot{x}_{i}\diagup R$ which can be resolved with respect to the
regular velocities as%
\begin{equation}
\dot{x}_{i}=\dot{x}_{0}\dfrac{p_{i}}{E},\ \ \ \ E=\sqrt{m^{2}+\sum
_{i=x,y,z}p_{i}^{2}}. \label{e}%
\end{equation}
Using (\ref{h}) and (\ref{hph1}) we obtain%
\begin{equation}
H_{phys}=0,\ \ \ \ B_{x_{0}}=\dfrac{\partial L}{\partial\dot{x}_{0}}%
=-m\dfrac{\dot{x}_{0}}{R}=-E, \label{hh1}%
\end{equation}
and so the \textquotedblleft physical sense\textquotedblright\ of $\left(
-B_{x_{0}}\right)  $ is that it is indeed the energy (\ref{e}), while the
\textquotedblleft physical\textquotedblright\ Hamiltonian is zero. Equations
of motion (\ref{q1})--(\ref{q2}) are%
\[
\dot{x}_{i}=\dot{x}_{0}\dfrac{p_{i}}{E},\ \ \ \ \ \dot{p}_{i}=\dfrac{\partial
B_{x_{0}}}{\partial x_{i}}\dot{x}_{0}=0,
\]
which coincide with the Lagrange equations following from (\ref{lp}). Note
that the velocity $\dot{x}_{0}$ is arbitrary here, and therefore we have one
\textquotedblleft gauge\textquotedblright\ degree of freedom.
\end{example}

\section{Nonabelian gauge theory interpretation}

We observe that (\ref{q3}) can be written in a more compact form using the
gauge theory notation. Let us introduce a \textquotedblleft$q^{\alpha}$-long
derivative\textquotedblright%
\begin{equation}
D_{\alpha}X=\dfrac{\partial X}{\partial q^{\alpha}}+\left\{  B_{\alpha
},X\right\}  _{phys}, \label{da}%
\end{equation}
where $X=X\left(  q^{A},p_{i}\right)  $ is a smooth scalar function on
$M_{phys}$. We also notice in (\ref{q3}) a \textquotedblleft$q^{\alpha}$-field
strength\textquotedblright\ $F_{\alpha\beta}\equiv F_{\alpha\beta}\left(
q^{A},p_{i}\right)  $ of the \textquotedblleft$q^{\alpha}$-gauge
fields\textquotedblright\ $B_{\alpha}$ on $M_{phys}$ defined by%
\begin{equation}
F_{\alpha\beta}=\dfrac{\partial B_{\beta}}{\partial q^{\alpha}}-\dfrac
{\partial B_{\alpha}}{\partial q^{\beta}}+\left\{  B_{\alpha},B_{\beta
}\right\}  _{phys}. \label{f}%
\end{equation}
Then the linear system of equations (\ref{q3}) for unresolved velocities can
written in a compact form%
\begin{equation}
\sum_{\beta=r+1}^{n}F_{\alpha\beta}\dfrac{dq^{\beta}}{dt}=D_{\alpha}%
H_{phys},\ \ \ \ \ \alpha=r+1,\ldots,n. \label{q3a}%
\end{equation}
The \textquotedblleft$q^{\alpha}$-field strength\textquotedblright%
\ $F_{\alpha\beta}$ is nonabelian due to the presence of the \textquotedblleft
physical\textquotedblright\ Poisson bracket in r.h.s. of (\ref{f}). It is
important to observe that, in distinct of the ordinary Yang-Mills theory, the
partial derivatives of $B_{\alpha}$ in (\ref{f}) are defined in the
$q^{\alpha}$-subspace $R^{n-r}$, while the \textquotedblleft
nonabelianity\textquotedblright\ (third term) is due to the Poisson bracket
(\ref{xyp}) in another symplectic subspace $Sp\left(  r,r\right)  $.

Note that the \textquotedblleft$q^{\alpha}$-long derivative\textquotedblright%
\ satisfies the Leibniz rule%
\[
D_{\alpha}\left\{  B_{\beta},B_{\gamma}\right\}  _{phys}=\left\{  D_{\alpha
}B_{\beta},B_{\gamma}\right\}  _{phys}+\left\{  B_{\beta},D_{\alpha}B_{\gamma
}\right\}  _{phys}%
\]
which is valid while acting on \textquotedblleft$q^{\alpha}$-gauge
fields\textquotedblright\ $B_{\alpha}$. The commutator of the
\textquotedblleft$q^{\alpha}$-long derivatives\textquotedblright\ is now equal
to the Poisson bracket with the \textquotedblleft$q^{\alpha}$-field
strength\textquotedblright%
\begin{equation}
\left(  D_{\alpha}D_{\beta}-D_{\beta}D_{\alpha}\right)  X=\left\{
F_{\alpha\beta},X\right\}  _{phys}. \label{ddf}%
\end{equation}

It follows from (\ref{ddf})%
\begin{equation}
D_{\alpha}D_{\beta}F_{\alpha\beta}=0. \label{ddf1}%
\end{equation}

Let us introduce the $B_{\alpha}$-transformation%
\begin{equation}
\mathbf{\delta}_{B_{\alpha}}X=\left\{  B_{\alpha},X\right\}  _{phys},
\label{db}%
\end{equation}
which satisfies%
\begin{align}
\left(  \mathbf{\delta}_{B_{\alpha}}\mathbf{\delta}_{B_{\beta}}-\mathbf{\delta
}_{B_{\beta}}\mathbf{\delta}_{B_{\alpha}}\right)  B_{\gamma}  &
=\mathbf{\delta}_{\left\{  B_{\alpha},B_{\beta}\right\}  _{phys}}B_{\gamma
},\label{dd1}\\
\mathbf{\delta}_{B_{\alpha}}F_{\beta\gamma}\left(  q^{A},p_{i}\right)   &
=\left(  D_{\gamma}D_{\beta}-D_{\beta}D_{\gamma}\right)  B_{\alpha
},\label{dd2}\\
\mathbf{\delta}_{B_{\alpha}}\left\{  B_{\beta},B_{\gamma}\right\}  _{phys}  &
=\left\{  \mathbf{\delta}_{B_{\alpha}}B_{\beta},B_{\gamma}\right\}
_{phys}+\left\{  B_{\beta},\mathbf{\delta}_{B_{\alpha}}B_{\gamma}\right\}
_{phys}. \label{dd3}%
\end{align}
This means that the \textquotedblleft$q^{\alpha}$-long
derivative\textquotedblright\ $D_{\alpha}$\ (\ref{da}) is in fact a
\textquotedblleft$q^{\alpha}$-covariant derivative\textquotedblright\ with
respect to the $B_{\alpha}$-transformation (\ref{db}). Indeed, observe that
\textquotedblleft$D_{\alpha}$ transforms as fields\textquotedblright%
\ (\ref{db}), which proves that it is really covariant (note the cyclic
permutations in both sides)
\begin{align}
&  \mathbf{\delta}_{B_{\alpha}}D_{\beta}B_{\gamma}+\mathbf{\delta}_{B_{\gamma
}}D_{\alpha}B_{\beta}+\mathbf{\delta}_{B_{\beta}}D_{\gamma}B_{\alpha
}\nonumber\\
&  =\left\{  B_{\alpha},D_{\beta}B_{\gamma}\right\}  _{phys}+\left\{
B_{\gamma},D_{\alpha}B_{\beta}\right\}  _{phys}+\left\{  B_{\beta},D_{\gamma
}B_{\alpha}\right\}  _{phys}.
\end{align}

The \textquotedblleft$q^{\alpha}$-Maxwell\textquotedblright\ equations of
motion for the \textquotedblleft$q^{\alpha}$-field strength\textquotedblright%
\ are%
\begin{align}
&  D_{\alpha}F_{\alpha\beta}=J_{\beta},\label{qm1}\\
&  D_{\alpha}F_{\beta\gamma}+D_{\gamma}F_{\alpha\beta}+D_{\beta}%
F_{\gamma\alpha}=0, \label{qm2}%
\end{align}
where $J_{\alpha}\equiv J_{\alpha}\left(  q^{A},p_{i}\right)  $ is a
\textquotedblleft$q^{\alpha}$-current\textquotedblright\ in $M_{phys}$ which
is a function of the initial Lagrangian (\ref{g}) and its derivatives up to
third order. Due to (\ref{ddf1}) the \textquotedblleft$q^{\alpha}%
$-current\textquotedblright\ $J_{\alpha}$ is conserved
\begin{equation}
D_{\alpha}J_{\alpha}=0.
\end{equation}
Thus, a singular Lagrangian system leads effectively to some special kind of
the nonabelian gauge\ theory in the direct product space $M_{phys}%
=R^{n-r}\times Sp\left(  r,r\right)  $. Here the \textquotedblleft
nonabelianity\textquotedblright\ (third term in (\ref{f})) appears not due to
a Lie algebra (as in the Yang-Mills theory), but \textquotedblleft
classically\textquotedblright, due to the Poisson bracket in the symplectic
subspace $Sp\left(  r,r\right)  $. The corresponding manifold can be
interpreted locally as a special kind of\ the degenerate Poisson manifold
(see, e.g., \cite{bue}).

The analogous Poisson type of \textquotedblleft
nonabelianity\textquotedblright\ (\ref{f}) appears in the $N\rightarrow\infty$
limit of Yang-Mills theory, and it is called the \textquotedblleft Poisson
gauge theory\textquotedblright\ \cite{flo/ili/tik}. In the $SU\left(
\infty\right)  $ Yang-Mills theory the group indices become surface
coordinates \cite{fai/fle/zac}, and it is connected with the Schild string
\cite{zac90}. The related algebra generalizations are called the continuum
graded Lie algebras \cite{sav/ver} (see, also, \cite{kon/sch2}). Here, because
of the direct product structure of the space $M_{phys}$, the similar
construction appears in (\ref{f}) (in another initial context), while the
\textquotedblleft long derivative\textquotedblright\ (\ref{da}),
\textquotedblleft gauge transformations\textquotedblright\ (\ref{db}) and the
analog of the Maxwell\ equations (\ref{qm1})--(\ref{qm2}) differ from the
\textquotedblleft Poisson gauge theory\textquotedblright\ \cite{flo/ili/tik}.

\section{\label{sect-class}Classification, gauge freedom and new brackets}

Next we can classify singular Lagrangian theories as follows:

\begin{enumerate}
\item \textit{Gaugeless theory}. The rank of the skew-symmetric matrix
$F_{\alpha\beta}$ is \textquotedblleft full\textquotedblright, i.e.
$\operatorname{rank}F_{\alpha\beta}=n-r$ and constant, and so the matrix
$F_{\alpha\beta}$ is invertible, and all the (degenerate) velocities
$v^{\alpha}$ can be found from the system of linear equations (\ref{q3}) (and
(\ref{q3a})) in a purely algebraic way.

\item \textit{Gauge theory}. The skew-symmetric matrix $F_{\alpha\beta}$ is
singular. If $\operatorname{rank}F_{\alpha\beta}=r_{F}<n-r$, then a singular
Lagrangian theory has $n-r-r_{F}$ gauge degrees of freedom. We can take them
arbitrary, which corresponds to the presence of some symmetries in the theory.
Note that the rank $r_{F}$ is even due to skew-symmetry of $F_{\alpha\beta}$.
\end{enumerate}

In the first case (gaugeless theory) one can resolve (\ref{q3a}) as follows%
\begin{equation}
v^{\beta}=\sum_{\alpha=r+1}^{n}\bar{F}^{\beta\alpha}D_{\alpha}H_{phys}%
,\label{vd}%
\end{equation}
where $\bar{F}^{\alpha\beta}$is the inverse matrix to $F_{\alpha\beta}$, i.e.
\begin{equation}
\sum_{\beta=r+1}^{n}F_{\alpha\beta}\bar{F}^{\beta\gamma}=\sum_{\beta=r+1}%
^{n}\bar{F}^{\gamma\beta}F_{\beta\alpha}=\delta_{\alpha}^{\gamma}.\label{ffff}%
\end{equation}
Substitute (\ref{vd}) in (\ref{q1})--(\ref{q2}) to present the system of
equations for a gaugeless degenerate Lagrangian theory in the Hamilton-like
form as follows%
\begin{align}
\dfrac{dq^{i}}{dt} &  =\left\{  q^{i},H_{phys}\right\}  _{nongauge}%
,\label{qh0}\\
\dfrac{dp_{i}}{dt} &  =\left\{  p_{i},H_{phys}\right\}  _{nongauge}%
,\label{ph0}%
\end{align}
where the \textquotedblleft nongauge\textquotedblright\ bracket is defined by%
\begin{equation}
\left\{  X,Y\right\}  _{nongauge}=\left\{  X,Y\right\}  _{phys}-\sum
_{\alpha=r+1}^{n}\sum_{\beta=r+1}^{n}D_{\alpha}X\cdot\bar{F}^{\alpha\beta
}\cdot D_{\beta}Y.\label{xyf}%
\end{equation}

Then the time evolution of any function of dynamical variables $X=X\left(
q^{A},p_{i}\right)  $ is also determined by the bracket (\ref{xyf}) as follows%
\begin{equation}
\dfrac{dX}{dt}=\left\{  X,H_{phys}\right\}  _{nongauge}. \label{dx}%
\end{equation}

The meaning of the new nongauge bracket (\ref{xyf}) (which appear naturally in
the Clairaut-type formulation \cite{dup2011}) is the same the meaning of the
ordinary Poisson bracket in the unconstrained Hamiltonian dynamics: it governs
the dynamics by the set of the first-order differential equations in the
Hamilton-like form (\ref{qh0})--(\ref{ph0}) and is responsible for the time
evolution of any dynamical variable (\ref{dx}). Also, the second term in the
new bracket (\ref{xyf}) has complicated coordinate dependence and is analogous
to that of the Dirac bracket \cite{dirac}. In the extended phase space the
both brackets coincide (see Appendix \ref{sect-a2}). On the other hand, the
appearance of the second term in (\ref{xyf}) can be treated as deformation of
the Poisson bracket, which can lead to another kind of generalized symplectic
geometry \cite{hit03}.

In the second case (gauge theory), with the singular matrix $F_{\alpha\beta}$
of rank $r_{F}$, we rearrange its rows and columns to obtain a nonsingular
$r_{F}\times r_{F}$ submatrix in the left upper corner. In such a way, the
first $r_{F}$ equations of the system of linear (under also rearranged
$v^{\beta}$) equations (\ref{q3a}) are independent. Then we express the
indices $\alpha$ and $\beta$ as pairs $\alpha=\left(  \alpha_{1},\alpha
_{2}\right)  $ and $\beta=\left(  \beta_{1},\beta_{2}\right)  $, where
$\alpha_{1}$ and $\beta_{1}$ denote the first $r_{F}$ rows and columns, while
$\alpha_{2}$ and $\beta_{2}$ denote the rest of $n-r-r_{F}$ rows and columns.
Correspondingly, we decompose the system (\ref{q3a}) as%
\begin{align}
\sum_{\beta_{1}=r+1}^{r+r_{F}}F_{\alpha_{1}\beta_{1}}v^{\beta_{1}}+\sum
_{\beta_{2}=r+r_{F}+1}^{n}F_{\alpha_{1}\beta_{2}}v^{\beta_{2}}  &
=D_{\alpha_{1}}H_{phys},\label{ff1}\\
\sum_{\beta_{1}=r+1}^{r+r_{F}}F_{\alpha_{2}\beta_{1}}v^{\beta_{1}}+\sum
_{\beta_{2}=r+r_{F}+1}^{n}F_{\alpha_{2}\beta_{2}}v^{\beta_{2}}  &
=D_{\alpha_{2}}H_{phys}. \label{ff2}%
\end{align}

Because the matrix $F_{\alpha_{1}\beta_{1}}$is nonsingular by construction, we
can find the first $r_{F}$ velocities%
\begin{equation}
v^{\beta_{1}}=\sum_{\alpha_{1}=r+1}^{r+r_{F}}\bar{F}^{\beta_{1}\alpha_{1}%
}D_{\alpha_{1}}H_{phys}-\sum_{\alpha_{1}=r+1}^{r+r_{F}}\bar{F}^{\beta
_{1}\alpha_{1}}F_{\alpha_{1}\beta_{2}}v^{\beta_{2}}, \label{v1}%
\end{equation}
where $\bar{F}^{\beta_{1}\alpha_{1}}$is the inverse of the nonsingular
$r_{F}\times r_{F}$ submatrix $F_{\alpha_{1}\beta_{1}}$satisfying (\ref{ffff}).

Then, since $\operatorname{rank}F_{\alpha\beta}=r_{F}$, the last $n-r-r_{F}$
equations (\ref{ff2}) are linear combinations of the first $r_{F}$ independent
ones (\ref{ff1}), which gives%
\begin{align}
F_{\alpha_{2}\beta_{1}} &  =\sum_{\alpha_{1}=r+1}^{r+r_{F}}\lambda_{\alpha
_{2}}^{\alpha_{1}}F_{\alpha_{1}\beta_{1}},\label{f1}\\
F_{\alpha_{2}\beta_{2}} &  =\sum_{\alpha_{1}=r+1}^{r+r_{F}}\lambda_{\alpha
_{2}}^{\alpha_{1}}F_{\alpha_{1}\beta_{2}},\label{f2}\\
D_{\alpha_{2}}H_{phys} &  =\sum_{\alpha_{1}=r+1}^{r+r_{F}}\lambda_{\alpha_{2}%
}^{\alpha_{1}}D_{\alpha_{1}}H_{phys},\label{dh}%
\end{align}
where $\lambda_{\alpha_{2}}^{\alpha_{1}}=\lambda_{\alpha_{2}}^{\alpha_{1}%
}\left(  q^{A},p_{i}\right)  $ are some $r_{F}\times\left(  n-r-r_{F}\right)
$ smooth functions. Using the relation (\ref{f1}) and invertibility of
$F_{\alpha_{1}\beta_{1}}$we eliminate the functions $\lambda_{\alpha_{2}%
}^{\alpha_{1}}$ by%
\begin{equation}
\lambda_{\alpha_{2}}^{\alpha_{1}}=\sum_{\alpha_{1}=r+1}^{r+r_{F}}\sum
_{\beta_{1}=r+1}^{r+r_{F}}F_{\alpha_{2}\beta_{1}}\bar{F}^{\beta_{1}\alpha_{1}%
}.
\end{equation}

This indicates that the gauge theory is fully determined by the first $r_{F}$
rows of the (rearranged) matrix $F_{\alpha\beta}$ and the first $r_{F}$
(rearranged) derivatives $D_{\alpha_{1}}H_{phys}$ only.

Next, we can make the unresolved $n-r-r_{F}$ velocities vanish
\begin{equation}
v^{\beta_{2}}=0\label{vb2}%
\end{equation}
by some \textquotedblleft gauge fixing\textquotedblright\ condition. Then
(\ref{v1}) becomes%
\begin{equation}
v^{\beta_{1}}=\sum_{\alpha_{1}=r+1}^{r+r_{F}}\bar{F}^{\beta_{1}\alpha_{1}%
}D_{\alpha_{1}}H_{phys}.\label{v2}%
\end{equation}
By analogy with (\ref{qh0})--(\ref{ph0}), in the gauge case we can also write
the system of equations for a singular Lagrangian theory in the Hamilton-like
form. Now we introduce another new (gauge) bracket
\begin{equation}
\left\{  X,Y\right\}  _{gauge}=\left\{  X,Y\right\}  _{phys}-\sum_{\alpha
_{1}=r+1}^{r+r_{F}}\sum_{\beta_{1}=r+1}^{r+r_{F}}D_{\alpha_{1}}X\cdot\bar
{F}^{\alpha_{1}\beta_{1}}\cdot D_{\beta_{1}}Y,\label{xyf1}%
\end{equation}

Then substituting (\ref{vb2})--(\ref{v2}) into (\ref{q1})--(\ref{q2}) and
using (\ref{xyf1}), we obtain%

\begin{align}
\dfrac{dq^{i}}{dt} &  =\left\{  q^{i},H_{phys}\right\}  _{gauge},\label{qqh}\\
\dfrac{dp_{i}}{dt} &  =\left\{  p_{i},H_{phys}\right\}  _{gauge}.\label{pph}%
\end{align}
Thus the gauge bracket (\ref{xyf1}) governs the time evolution in the gauge
case%
\begin{equation}
\dfrac{dX}{dt}=\left\{  X,H_{phys}\right\}  _{gauge}.\label{dxf1}%
\end{equation}

Note that the brackets (\ref{xyf}) and (\ref{xyf1}) are antisymmetric and
satisfy the Jacobi identity. Therefore, the standard quantization scheme is
applicable here (see, e.g. \cite{gre/rei}). The difference is the fact that
only the canonical (regular) dynamic variables $\left(  q^{i},p_{i}\right)  $
should be quantized, while the degenerate coordinates can be trated as some
continuous parameters.

It is worthwhile to consider the \textit{limit case}, when $r_{F}=0$, i.e.
\begin{equation}
F_{\alpha\beta}=0\label{f0}%
\end{equation}
identically, which can mean that $B_{\alpha}=0$, so the Lagrangian can be
independent of the degenerate velocities $v^{\alpha}$. It follows from
(\ref{q3}) that%
\begin{equation}
D_{\alpha}H_{phys}=\dfrac{\partial H_{phys}}{\partial q^{\alpha}%
}=0,\label{hqq}%
\end{equation}
which leads to \textit{the \textquotedblleft independence\textquotedblright%
\ statement}: the \textquotedblleft physical\textquotedblright\ Hamiltonian
$H_{phys}$ does not depend on the degenerate coordinates $q^{\alpha}$, iff the
Lagrangian does not depend on the velocities $v^{\alpha}$. In the limit case,
the both brackets (\ref{xyf}) and (\ref{xyf1}) coincide with the Poisson
bracket in the reduced (\textquotedblleft physical\textquotedblright)\ phase
space $\left\{  \ ,\ \right\}  _{nongauge,gauge}=\left\{  \ ,\ \right\}
_{phys}$.

\begin{example}
\textrm{(Christ-Lee model \cite{chr/lee})} The Lagrangian of $SU\left(
2\right)  $ Yang-Mills theory in $0+1$ dimensions is (in our notation)%
\begin{equation}
L\left(  x_{i},y_{\alpha},v_{i}\right)  =\dfrac{1}{2}\sum_{i=1,2,3}\left(
v_{i}-\sum_{j,\alpha=1,2,3}\varepsilon_{ij\alpha}x_{j}y_{\alpha}\right)
^{2}-U\left(  x^{2}\right)  , \label{lcl}%
\end{equation}
where $i,\alpha=1,2,3$, $x^{2}=\sum_{i}x_{i}^{2}$, $v_{i}=\dot{x}_{i}$ and
$\varepsilon_{ijk}$ is the Levi-Civita symbol. Because (\ref{lcl}) is
independent of degenerate velocities $\dot{y}_{\alpha}$, all $B_{\alpha
}\overset{\mathrm{(\ref{h})}}{=}0$, and therefore $F_{\alpha\beta}%
\overset{\mathrm{(\ref{f})}}{=}0$, we have the limit gauge case of the above
classification. The corresponding Clairaut equation (\ref{cl}) for
$H=H^{\mathrm{Cl}}\left(  x_{i},y_{\alpha},\lambda_{i},\lambda_{\alpha
}\right)  $ has the form%
\begin{equation}
H=\sum_{i=1,2,3}\lambda_{i}H_{\lambda_{i}}^{\prime}+\sum_{\alpha=1,2,3}%
\lambda_{\alpha}H_{\lambda_{\alpha}}^{\prime}-\dfrac{1}{2}\sum_{i=1,2,3}%
\left(  H_{\lambda_{i}}^{\prime}-\sum_{j,\alpha=1,2,3}\varepsilon_{ij\alpha
}x_{j}y_{\alpha}\right)  ^{2}+U\left(  x^{2}\right)  . \label{hu}%
\end{equation}
We show manifestly, how to obtain the envelope solution for regular variables
and general solution for degenerate variables. Its general solution is%
\begin{equation}
H_{gen}=\sum_{i=1,2,3}\lambda_{i}c_{i}+\sum_{\alpha=1,2,3}\lambda_{\alpha
}c_{\alpha}-\dfrac{1}{2}\sum_{i=1,2,3}\left(  c_{i}-\sum_{j,\alpha
=1,2,3}\varepsilon_{ij\alpha}x_{j}y_{\alpha}\right)  ^{2}+U\left(
x^{2}\right)  , \label{hgen}%
\end{equation}
where $c_{i}$, $c_{\alpha}$ are arbitrary functions of coordinates. Recall
that $q^{A}$ are passive variables under the Legendre transform. We
differentiate (\ref{hgen}) by $c_{i}$, $c_{\alpha}$
\begin{align}
\dfrac{\partial H_{gen}}{\partial c_{i}}  &  =\lambda_{i}-\left(  c_{i}%
-\sum_{j,\alpha=1,2,3}\varepsilon_{ij\alpha}x_{j}y_{\alpha}\right)
,\label{hg1}\\
\dfrac{\partial H_{gen}}{\partial c_{\alpha}}  &  =\lambda_{\alpha},
\label{hg2}%
\end{align}
and observe that only the first relation (\ref{hg1}) can be resolved with
respect to $c_{i}$, and therefore can lead to the envelope solution, while
other $c_{\alpha}$ cannot be resolved, and therefore we consider only general
solution of the Clairaut equation. So we can exclude half of the constants
using (\ref{hg1}) (with the substitution $\lambda_{i}\overset
{\mathrm{(\ref{pp})}}{\rightarrow}p_{i}$) and get the mixed solution
(\ref{hm}) to the Clairaut equation (\ref{hu}) as%
\begin{equation}
H_{mix}^{\mathrm{Cl}}\left(  x_{i},y_{\alpha},p_{i},\lambda_{\alpha}%
,c_{\alpha}\right)  =\dfrac{1}{2}\sum_{i=1,2,3}p_{i}^{2}+\sum_{i,j,\alpha
=1,2,3}\varepsilon_{ij\alpha}p_{i}x_{j}y_{\alpha}+\sum_{\alpha=1,2,3}%
\lambda_{\alpha}c_{\alpha}+U\left(  x^{2}\right)  .
\end{equation}
Using (\ref{hph}), we obtain the \textquotedblleft physical\textquotedblright%
\ Hamiltonian%
\begin{equation}
H_{phys}\left(  x_{i},y_{\alpha},p_{i}\right)  =\dfrac{1}{2}\sum
_{i=1,2,3}p_{i}^{2}+\sum_{i,j,\alpha=1,2,3}\varepsilon_{ij\alpha}p_{i}%
x_{j}y_{\alpha}+U\left(  x^{2}\right)  . \label{hphys}%
\end{equation}

\ From the other side, the Hessian of (\ref{lcl}) has rank $3$, and we choose
$x_{i}$, $v_{i}$ and $y_{\alpha}$ to be regular and degenerate variables
respectively. The degenerate velocities $v_{\alpha}=\dot{y}_{\alpha}$ cannot
be defined from (\ref{q3}) at all, they are arbitrary, and the first integrals
(\ref{q3}), (\ref{hqq}) of the system (\ref{q1})--(\ref{q2}) become (also in
accordance to the independence statement)%
\begin{equation}
\dfrac{\partial H_{phys}\left(  x_{i},y_{\alpha},p_{i}\right)  }{\partial
y_{\alpha}}=\sum_{i,j=1,2,3}\varepsilon_{ij\alpha}p_{i}x_{j}=0. \label{hya}%
\end{equation}
The preservation in time (\ref{dxf1}) of (\ref{hya}) is fulfilled identically
due to the antisymmetry properties of the Levi-Civita symbols. It is clear
that only 2 equations from 3 of (\ref{hya}) are independent, so we choose
$p_{1}x_{2}=p_{2}x_{1}$, $p_{1}x_{3}=p_{3}x_{1}$ and insert in (\ref{hphys})
to get%
\begin{equation}
\tilde{H}_{phys}=\dfrac{1}{2}p_{1}^{2}\dfrac{x^{2}}{x_{1}^{2}}+U\left(
x^{2}\right)  .
\end{equation}
The transformation $\tilde{p}=p_{1}\sqrt{x^{2}}\diagup x_{1}$, $\tilde
{x}=\sqrt{x^{2}}$ gives the well-known result \cite{chr/lee,gog/khv/mla/pav}%
\begin{equation}
\tilde{H}_{phys}=\dfrac{1}{2}\tilde{p}^{2}+U\left(  \tilde{x}\right)  .
\end{equation}

\end{example}

\section{\label{sec-many}Singular Lagrangian systems and many-time dynamics}

The many-time classical dynamics and its connection with constrained systems
were studied in \cite{lon/lus/pon,kom78} as a generalization of some
relativistic two-particle models \cite{dro-vin69}. We consider this connection
from a different viewpoint, that is in the Clairaut-type approach
\cite{dup2011}. Recall that the Hamiltonian-Clairaut dynamics (\ref{q1}%
)--(\ref{q3}) of a Lagrangian singular system (\ref{leq}) is governed by the
\textquotedblleft physical\textquotedblright\ Hamiltonian function $H_{phys}$
and $\left(  n-r\right)  $ \textquotedblleft$q^{\alpha}$-gauge
fields\textquotedblright\ $B_{\alpha}$ defined on the direct product space
$R^{n-r}\times Sp\left(  r,r\right)  $. Let us treat the degenerate
coordinates $q^{\alpha}\in R^{n-r}$ as $\left(  n-r\right)  $ additional
\textquotedblleft time\textquotedblright\ variables together with $\left(
n-r\right)  $ corresponding \textquotedblleft Hamiltonians\textquotedblright%
\ $-B_{\alpha}\left(  q^{\alpha}|q^{i},p_{i}\right)  $, $\alpha=r+1,\ldots,n$
(see (\ref{da})). Indeed, let us introduce $\left(  n-r+1\right)  $
generalized \textquotedblleft times\textquotedblright\ $\mathsf{t}^{\mu}$ and
the corresponding \textquotedblleft many-time Hamiltonians\textquotedblright%
\ $\mathsf{H}_{\mu}\left(  \mathsf{t}^{\mu}|q^{i},p_{i}\right)  $,
$\mu=0,\ldots n-r$ defined by%
\begin{align}
\mathsf{t}^{0} &  =t,\ \ \ \mathsf{H}_{0}\left(  \mathsf{t}^{\alpha}%
|q^{i},p_{i}\right)  =H_{phys}\left(  q^{\alpha}|q^{i},p_{i}\right)
,\ \ \ \mu=0,\label{th1}\\
\mathsf{t}^{\mu} &  =q^{\mu},\ \ \ \mathsf{H}_{\mu}\left(  \mathsf{t}^{\mu
}|q^{i},p_{i}\right)  =-B_{r+\mu}\left(  q^{r+\mu}|q^{i},p_{i}\right)
,\ \ \ \mu=1,\ldots,n-r.\label{th2}%
\end{align}

Then the equations (\ref{q1})--(\ref{q2}) can be presented in the differential
form%
\begin{align}
dq^{i}  &  =\sum_{\mu=0}^{n-r}\left\{  q^{i},\mathsf{H}_{\mu}\right\}
_{phys}d\mathsf{t}^{\mu},\label{qh1}\\
dp_{i}  &  =\sum_{\mu=0}^{n-r}\left\{  p_{i},\mathsf{H}_{\mu}\right\}
_{phys}d\mathsf{t}^{\mu}, \label{qh2}%
\end{align}
where $\left\{  \ ,\ \right\}  _{phys}$ is defined in (\ref{xyp}). The linear
algebraic system of equations (\ref{q3}) for the degenerate velocities then
becomes%
\begin{equation}
\sum_{\mu=0}^{n-r}\mathsf{G}_{\mu\nu}d\mathsf{t}^{\mu}=0, \label{psh}%
\end{equation}
where%
\begin{equation}
\mathsf{G}_{\mu\nu}=\dfrac{\partial\mathsf{H}_{\mu}}{\partial\mathsf{t}^{\nu}%
}-\dfrac{\partial\mathsf{H}_{\nu}}{\partial\mathsf{t}^{\mu}}+\left\{
\mathsf{H}_{\mu},\mathsf{H}_{\nu}\right\}  _{phys}. \label{hhh}%
\end{equation}

It follows from (\ref{psh}) that the one-form $\omega=p_{i}dq^{i}%
-\mathsf{H}_{\mu}d\mathsf{t}^{\mu}$ is closed%
\begin{equation}
d\omega=\dfrac{1}{2}\sum_{\mu=0}^{n-r}\sum_{\nu=0}^{n-r}\mathsf{G}_{\mu\nu
}d\mathsf{t}^{\mu}\wedge d\mathsf{t}^{\nu}=0,\label{ptt}%
\end{equation}
which agrees with the action principle for multi-time classical dynamics
\cite{dom/lon/gom/pon}. The corresponding set of the Hamilton-Jacobi equations
for action $S\left(  q^{\alpha}|q^{i},p_{i}\right)  \longmapsto\mathsf{S}%
\left(  \mathsf{t}^{\mu}|q^{i},p_{i}\right)  $ is%
\begin{equation}
\dfrac{\partial\mathsf{S}}{\partial\mathsf{t}^{\mu}}+\mathsf{H}_{\mu}\left(
\mathsf{t}^{\mu}|q^{i},\dfrac{\partial\mathsf{S}}{\partial q^{i}}\right)
=0.\label{st}%
\end{equation}

Therefore, we come to the conclusion that any singular Lagrangian theory (in
the\ Clairaut-type formulation \cite{dup_belg2009,dup2011}) is equivalent to
the many-time classical dynamics \cite{dom/lon/gom/pon,lon/lus/pon}: the
equations of motion are (\ref{qh1})--(\ref{qh2}) (which coincide with
(\ref{q1})--(\ref{q2})), and the integrability condition is (\ref{psh}) which
coincides with the system of linear algebraic equations for unresolved
velocities (\ref{q3}) by construction.

\section{Conclusions}

We have described Hamilton-like evolution of singular Lagrangian systems using
$n-r+1$ functions $H_{phys}\left(  q^{\alpha}|q^{i},p_{i}\right)  $ and
$B_{\alpha}\left(  q^{\alpha}|q^{i},p_{i}\right)  $ on the direct product
space $R^{n-r}\times Sp\left(  r,r\right)  $. This is done by means of the
generalized Legendre-Clairaut transform, that is by solving the corresponding
multidimensional Clairaut equation without introducing the Lagrange
multipliers. All variables are set as regular or degenerate according to the
rank of the Hessian matrix of Lagrangian. We consider the reduced
\textquotedblleft physical\textquotedblright\ phase space formed by the
regular coordinates $q^{i}$ and momenta $p_{i}$ only, while degenerate
coordinates $q^{\alpha}$ play a role of parameters. There are two reasons, why
the degenerate momenta $\lambda_{\alpha}$ corresponding to $q^{\alpha}$ need
not be considered in the Clairaut-type formulation:

1) \textit{the mathematical reason}: there is no possibility to find the
degenerate velocities $v^{\alpha}$, as can be done for the regular velocities
$v^{i}$ in (\ref{pp}), and the \textquotedblleft
pre-Hamiltonian\textquotedblright\ (\ref{g}) has no extremum in degenerate directions;

2) \textit{the physical reason}: momentum is a \textquotedblleft measure of
movement\textquotedblright, but in \textquotedblleft
degenerate\textquotedblright\ directions there is \textit{no dynamics}, hence
--- no reason to introduce the corresponding \textquotedblleft
physical\textquotedblright\ momenta at all.

Note that some possibilities to avoid constraints were considered in a
different context in \cite{der2005,ran08} and for special forms of the
Lagrangian in \cite{git/tyu1}.

The Hamilton-like form of the equations of motion (\ref{qh0})--(\ref{ph0}) is
achieved by introducing new brackets (\ref{xyf}) and (\ref{xyf1}) which are
responsible for the time evolution. They are antisymmetric and satisfy the
Jacobi identity. Therefore quantization of such brackets can be done by the
standard methods \cite{gre/rei}, but only for the regular variables, while the
degenerate variables can be considered as some continuous parameters.

In the \textquotedblleft nonphysical\textquotedblright\ coordinate subspace,
we formulate some kind of nonabelian gauge theory, such that \textquotedblleft
nonabelianity\textquotedblright\ appears due to the Poisson bracket in the
physical phase space (\ref{f}). This makes it similar to the Poisson gauge
theory \cite{flo/ili/tik}, but do not coincide with the latter.

Finally, we show that, in general, a singular Lagrangian system in the
Clairaut-type formulation \cite{dup_belg2009,dup2011} is equivalent to the
many-time classical dynamics.

\section*{Acknowledgements}

The author is grateful to G. A. Goldin and J. Lebowitz for kind hospitality at
the Rutgers University, where this work has been finalized, and to the
Fulbright Scholar Program for financial support, also he would like to express
deep thankfulness to Jim Stasheff for careful reading the final version and
making many corrections and important remarks.

The author would like to thank V.~P.~Aku\-lov, A.~V.~Anto\-nyuk, H.~Arodz,
Yu.~A.~Berezhnoj, V.~Berezovoj, Yu.~Bespalov, Yu.~Bolotin, L.~Bonora,
I.~H.~Brevik, B.~Broda, B.~Burgstaller, R.~Casalbuoni, M.~Dab\-row\-ski,
V.~K.~Dubovoy, M.~Du\-dek, P.~Etingof, V.~Gershun, M.~Ger\-sten\-haber,
D.~M.~Gitman, S.~Goldstein, D.~Grumiller, J.~Grabowski, H.~Grosse,
U.~G\"{u}nter, R.~Jackiw, J.~Grabowski, H.~Jones, R.~Khuri, V.~D.~Khodusov,
A.~T.~Kotvytskiy, M.~Krivoruchenko, G.~Ch.~Kurinnoj, S. Kuzhel, M.~Lapidus,
J.~Lukierski, L.~Lusanna, P.~Mahnke, N.~Merenkov, A.~A.~Migdal, M.~Mulase,
B.~V.~Novikov, A.~Nurmagambetov, L.~A.~Pastur, P.~Orland, S.~A.~Ovsienko,
M.~Pavlov, S.~V.~Peletminskij, D.~Polyakov, M.~Schli\-chen\-maier,
A.~S.~Schwarz, B.~Shapiro, M.~Shifman, V.~Shtelen, W.~Siegel, V.~A.~Soroka,
K.~S.~Stelle, Yu.~P.~Stepanovsky, M.~Tonin, W.~M.~Tulczyjew, R.~Umble,
P.~Urbanski, A.~Vainstein, M.~Voloshin, A.~Voronov, K.~Wali, M.~Walker,
A.~A.~Yantzevich, C.~Zachos, A.~A.~Zheltukhin, M.~Znojil and B.~Zwiebach for
fruitful discussions.

\appendix\setcounter{section}{0}

\section{Multidimensional Clairaut equation\label{sec-clair}}

The multidimensional Clairaut equation for a function $y=y(x_{i})$ of $n$
variables $x_{i}$ is \cite{izu94,ale/vin/lyc}%
\begin{equation}
y=\sum_{j=1}^{n}x_{j}y_{x_{j}}^{\prime}-f(y_{x_{i}}^{\prime}), \label{y}%
\end{equation}
where prime denotes a partial differentiation by subscript and $f$ is a smooth
function of $n$ arguments. To find and classify solutions of (\ref{y}), we
need to find first derivatives $y_{x_{i}}^{\prime}$ in some way, and then
substitute them back to (\ref{y}). We differentiate the Clairaut equation
(\ref{y}) by $x_{j}$ and obtain $n$ equations%
\begin{equation}
\sum_{i=1}^{n}y_{x_{i}x_{j}}^{\prime\prime}(x_{i}-f_{y_{x_{i}}^{\prime}%
}^{\prime})=0. \label{yx}%
\end{equation}
The classification follows from the ways the factors in (\ref{yx}) can be set
to zero. Here, for our physical applications, it is sufficient to suppose that
ranks of Hessians of $y$ and $f$ are equal%
\begin{equation}
\operatorname*{rank}y_{x_{i}x_{j}}^{\prime\prime}=\operatorname*{rank}%
f_{y_{x_{i}}^{\prime}y_{x_{j}}^{\prime}}^{\prime\prime}=r.
\end{equation}
This means that in each equation from (\ref{yx}) either the first or the
second multiplier is zero, but it is not necessary to vanish both of them. The
first multiplier can be set to zero without any additional assumptions. So we have

1) \textit{The general solution.} It is defined by the condition%
\begin{equation}
y_{x_{i}x_{j}}^{\prime\prime}=0. \label{yy}%
\end{equation}
After one integration we find $y_{x_{i}}^{\prime}=c_{i}$ and substitution them
to (\ref{y}) and obtain%
\begin{equation}
y_{gen}=\sum_{j=1}^{n}x_{j}c_{j}-f(c_{i}), \label{ygen}%
\end{equation}
where $c_{i}$ are $n$ constants.

All second multipliers in (\ref{yx}) can be zero for $i=1,\ldots,n$, but this
will give a solution, if we can resolve them under $y_{x_{i}}^{\prime}$. It
may be possible, if the rank of Hessians $f$ is full, i.e. $r=n$. In this case
we obtain

2) \textit{The envelope solution.} It is defined by
\begin{equation}
x_{i}=f_{y_{x_{i}}^{\prime}}^{\prime}. \label{xf}%
\end{equation}
We resolve (\ref{xf}) under derivatives as $y_{x_{i}}^{\prime}=C_{i}\left(
x_{j}\right)  $ and get%
\begin{equation}
y_{env}=\sum_{i=1}^{n}x_{i}C_{i}\left(  x_{j}\right)  -f(C_{i}\left(
x_{j}\right)  ), \label{yenv}%
\end{equation}
where $C_{i}\left(  x_{j}\right)  $ are $n$ smooth functions of $n$ arguments.

In the intermediate case, we can use the envelope solution (\ref{yenv}) for
first $s$ variables, while the general solution (\ref{ygen}) for other $n-s$
variables, and obtain

3) \textit{The }$s$\textit{-mixed solution}, as follows%
\begin{equation}
y_{mix}^{\left(  s\right)  }=\sum_{j=1}^{s}x_{j}C_{j}\left(  x_{j}\right)
+\sum_{j=s+1}^{n}x_{j}c_{j}-f(C_{1}\left(  x_{j}\right)  ,\ldots,C_{s}\left(
x_{j}\right)  ,c_{s+1},\ldots,c_{n}). \label{ymix}%
\end{equation}

If the rank $r$ of Hessians $f$ is not full and a nonsingular minor of the
rank $r$ is in upper left corner, then we can resolve first $r$ relations
(\ref{xf}) only, and so $s\leq r$. In our physical applications we use the
limited case $s=r$.

\textbf{Example. }Let $f\left(  z_{i}\right)  =z_{1}^{2}+z_{2}^{2}+z_{3}$,
then the Clairaut equation for $y=y\left(  x_{1},x_{2,}x_{3}\right)  $ is%
\begin{equation}
y=x_{1}y_{x_{1}}^{\prime}+x_{2}y_{x_{2}}^{\prime}+x_{3}y_{x_{3}}^{\prime
}-\left(  y_{x_{1}}^{\prime}\right)  ^{2}-\left(  y_{x_{2}}^{\prime}\right)
^{2}-y_{x_{3}}^{\prime},
\end{equation}
and we have $n=3$ and $r=2$. The general solution can be found from (\ref{yy})
by one integration and using (\ref{ygen})%
\begin{equation}
y_{gen}=c_{1}\left(  x_{1}-c_{1}\right)  +c_{2}\left(  x_{2}-c_{2}\right)
+c_{3}\left(  x_{3}-1\right)  ,
\end{equation}
where $c_{i}$ are constants.

Since $r=2$, we can resolve only $2$ relations from (\ref{xf}) by $y_{x_{1}%
}^{\prime}=\frac{x_{1}}{2}$, $y_{x_{2}}^{\prime}=\frac{x_{2}}{2}$. So there is
no envelope solution (for all variables), but we have several mixed solutions
corresponding to $s=1,2$:%
\begin{align}
y_{mix}^{\left(  1\right)  }  &  =\left\{
\begin{array}
[c]{c}%
\dfrac{x_{1}^{2}}{4}+c_{2}\left(  x_{2}-c_{2}\right)  +c_{3}\left(
x_{3}-1\right)  ,\\
c_{1}\left(  x_{1}-c_{1}\right)  +\dfrac{x_{2}^{2}}{4}+c_{3}\left(
x_{3}-1\right)  ,
\end{array}
\right. \\
y_{mix}^{\left(  2\right)  }  &  =\dfrac{x_{1}^{2}}{4}+\dfrac{x_{2}^{2}}%
{4}+c_{3}\left(  x_{3}-1\right)  .
\end{align}

The case $f\left(  z_{i}\right)  =z_{1}^{2}+z_{2}^{2}$ can be obtained from
the above formulas by putting $x_{3}=c_{3}=0$, while $y_{mix}^{\left(
2\right)  }$ becomes the envelope solution $y_{env}=\tfrac{x_{1}^{2}}%
{4}+\tfrac{x_{2}^{2}}{4}$.

\section{Correspondence with the Dirac approach\label{sect-a2}}

Here we explain the reason of appearence of constraints in the theories with
the degenerate Lagrangians: introducing into the theory \textit{additional}
dynamical variables (because the Hamilton-like form of the equations of motion
can be achieved \textit{without} them in the presented approach), that is
momenta which correspond to the \textquotedblleft degenerate\textquotedblright%
\ velocities. The connection of the Clairaut-type formulation with the Dirac
approach can be made by interpretation of the variables $\lambda_{\alpha}$
entering to the general solution of the Clairaut equation as the
\textquotedblleft physical\textquotedblright\ degenerate momenta $p_{\alpha}$
using for them the same expression through the Lagrangian as (\ref{pp})
\begin{equation}
\lambda_{\alpha}=p_{\alpha}=\dfrac{\partial L\left(  q^{A},v^{A}\right)
}{\partial v^{\alpha}}.\label{pp1}%
\end{equation}
Then we obtain the primary Dirac constraints (in the resolved form and our
notation (\ref{h}))%
\begin{equation}
\mathbf{\Phi}_{\alpha}\left(  q^{A},p_{A}\right)  =p_{\alpha}-B_{\alpha
}=0,\ \ \ \alpha=r+1,\ldots n,\label{fp}%
\end{equation}
which are defined now on the full phase space $\mathsf{T}^{\ast}M$. Using
(\ref{hph}) and (\ref{hph1}), we can arrive at the complete Hamiltonian of the
first-order formulation \cite{git/tyu0} (corresponding to the total Dirac
Hamiltonian \cite{dirac})%
\begin{align}
H_{T}\left(  q^{A},p_{A},v^{\alpha}\right)   &  =\left.  H_{mix}^{\mathrm{Cl}%
}\left(  q^{A},p_{i},\lambda_{\alpha},v^{\alpha}\right)  \right\vert
_{\lambda_{\alpha}=p_{\alpha}}\nonumber\\
&  =H_{phys}\left(  q^{A},p_{i}\right)  +\sum_{\alpha=r+1}^{n}v^{\alpha
}\mathbf{\Phi}_{\alpha}\left(  q^{A},p_{A}\right)  ,
\end{align}
which is equal to the mixed\ Hamilton-Clairaut function (\ref{hm}) with the
substitution (\ref{pp1}) and use of (\ref{fp}). Then the Hamilton-Clairaut
system of equations (\ref{q1})--(\ref{q2}) coincides with the Hamilton system
in the first-order formulation \cite{git/tyu0}%
\begin{equation}
\dot{q}^{A}=\left\{  q^{A},H_{T}\right\}  _{full},\ \ \ \ \dot{p}_{A}=\left\{
p_{A},H_{T}\right\}  _{full},\ \ \ \mathbf{\Phi}_{\alpha}=0,\label{qp}%
\end{equation}
and (\ref{q3}) gives the second stage equations of the Dirac approach%
\begin{equation}
\left\{  \mathbf{\Phi}_{\alpha},H_{T}\right\}  _{full}=\left\{  \mathbf{\Phi
}_{\alpha},H_{phys}\right\}  _{full}+\sum_{\beta=r+1}^{n}\left\{
\mathbf{\Phi}_{\alpha},\mathbf{\Phi}_{\beta}\right\}  _{full}v^{\beta
}=0,\label{ff}%
\end{equation}
where%
\begin{equation}
\left\{  X,Y\right\}  _{full}=\sum_{A=1}^{n}\left(  \frac{\partial X}{\partial
q^{A}}\frac{\partial Y}{\partial p_{A}}-\frac{\partial Y}{\partial q^{A}}%
\frac{\partial X}{\partial p_{A}}\right)
\end{equation}
is the (\textit{full})\textit{\ }Poisson bracket on the whole phase space
$\mathsf{T}^{\ast}M$. Note that%
\begin{align}
F_{\alpha\beta} &  =\left\{  \mathbf{\Phi}_{\alpha},\mathbf{\Phi}_{\beta
}\right\}  _{full},\label{fab}\\
D_{\alpha}H_{phys} &  =\left\{  \mathbf{\Phi}_{\alpha},H_{phys}\right\}
_{full}.\label{dhf}%
\end{align}

It is important that the introduced new brackets (\ref{xyf}) and (\ref{xyf1})
become the Dirac bracket \cite{dirac}. Moreover, our cases 2) and 1) of
Section \ref{sect-class} work as counterparts of the first and the second
class constraints in the Dirac classification \cite{dirac}, respectively. The
limit case with zero \textquotedblleft$q^{\alpha}$-field
strength\textquotedblright\ $F_{\alpha\beta}=0$ (\ref{f0}) (see (\ref{fab}))
corresponds to the Abelian constraints \cite{gog/khv/per,lor05}.

\end{document}